\documentclass[sigconf,screen]{acmart}
\usepackage{pifont}

\usepackage{booktabs}
\usepackage{multirow}
\usepackage{makecell}
\usepackage{hyperref}
\usepackage{listings}
\usepackage{xspace}
\usepackage{subcaption}
\usepackage{graphicx}

\usepackage{enumitem}
\usepackage{xcolor}
\usepackage{pifont}
\usepackage[most]{tcolorbox}

\newtcolorbox{rqbox}{
  colback=gray!15,
  colframe=blue!70!black,
  boxrule=0pt,
  left=6pt,
  right=6pt,
  top=6pt,
  bottom=6pt,
  borderline west={3pt}{0pt}{blue!70!black},
  borderline east={3pt}{0pt}{blue!70!black},
}

\newcommand{\tool}{\textsc{KQFuzz}\xspace}
\newcommand{\F}{Figure}
\newcommand{\E}{Eqn.}

\newcommand{\parh}[1]{\smallskip\noindent\textbf{#1}}

\definecolor{codegreen}{rgb}{0,0.6,0}
\definecolor{codegray}{rgb}{0.5,0.5,0.5}
\definecolor{codepurple}{rgb}{0.58,0,0.82}
\definecolor{backcolour}{rgb}{0.95,0.95,0.92}

\lstdefinestyle{mystyle}{
  backgroundcolor=\color{backcolour}, commentstyle=\color{codegreen},
  keywordstyle=\color{magenta},
  numberstyle=\tiny\color{codegray},
  stringstyle=\color{codepurple},
  basicstyle=\ttfamily\footnotesize,
  breakatwhitespace=false,         
  breaklines=true,                 
  captionpos=b,                    
  keepspaces=true,                 
  numbers=left,                    
  numbersep=5pt,                  
  showspaces=false,                
  showstringspaces=false,
  showtabs=false,                  
  tabsize=2
}

\AtBeginDocument{%
  }

\copyrightyear{2026}
\acmYear{2026}
\acmConference[ASE '26]
  {41st IEEE/ACM International Conference on Automated Software Engineering}
  {October 12--16, 2026}
  {Munich, Germany}

\usepackage[utf8]{inputenc}
\usepackage{xcolor}
\usepackage{listings}
\usepackage{caption}
\usepackage{subcaption} 
\usepackage{courier}

\definecolor{commentdarkgreen}{RGB}{0, 150, 0}
\definecolor{kwpurple}{RGB}{127, 0, 85}
\definecolor{gateblue}{RGB}{20, 60, 140}
\definecolor{methoddark}{RGB}{30, 140, 170}
\definecolor{strred}{RGB}{180, 0, 0}
\definecolor{numorange}{RGB}{160, 80, 0}
\definecolor{numgray}{RGB}{120, 120, 120}

\lstdefinelanguage{QuantumPy}{
    language=Python,
    morekeywords={as, name, from, import}, 
    keywordstyle=\color{kwpurple}\bfseries,
    morekeywords=[2]{QuantumCircuit, ECRGate, CHGate, ZGate, RCCXGate, iSwapGate, SdgGate, CXGate, Parameter, ForLoopOp, is_native_neutral_atom_gate, UniformSuperpositionGate, Clifford, synth_clifford_depth_lnn},
    keywordstyle=[2]\color{gateblue}\bfseries,
    morekeywords=[3]{append, qubits, qargs, cargs, h, ccx, measure_all, grover_operator, rx, assign_parameters, device, Hamiltonian, PauliZ, PauliX, sign_expand, qnode, RX, RY, CNOT, expval},
    keywordstyle=[3]\color{methoddark}, 
    sensitive=true,
    comment=[l]{\#},
    commentstyle=\color{commentdarkgreen}\ttfamily, 
    stringstyle=\color{strred},
    morestring=[b]',
    morestring=[b]",
}

\lstdefinestyle{QuantumPyBordered}{
    language=QuantumPy,
    frame=tlbr,
    framesep=5pt,
    framerule=0.1pt,
    framextopmargin=2pt,
    framexbottommargin=2pt,
    backgroundcolor=\color{gray!5},
}

\DeclareCaptionFormat{myformat}{\textbf{#1#2} #3}
\begin{document}

\title{\tool: Knowledge-Guided Fuzzing for Quantum Libraries via Large Language Models}

\author{Fuyuan Xia}
\affiliation{%
  \institution{Shanghai Jiao Tong University}
  \city{Shanghai}
  \country{China}
}
\email{fuyuanxia@sjtu.edu.cn}

\author{Qixin Zhang}
\affiliation{%
  \institution{Nanyang Technological University}
  \city{Singapore}
  \country{Singapore}
}
\email{qixin.zhang@ntu.edu.sg}

\author{Chenhao Ying}
\affiliation{%
  \institution{Shanghai Jiao Tong University}
  \city{Shanghai}
  \country{China}
}
\email{yingchenhao@sjtu.edu.cn}

\author{Haojin Zhu}
\affiliation{%
  \institution{Shanghai Jiao Tong University}
  \city{Shanghai}
  \country{China}
}
\email{zhu-hj@sjtu.edu.cn}

\author{Shuai Wang}
\affiliation{%
  \institution{Hong Kong University of Science and Technology}
  \city{Hong Kong}
  \country{China}
}
\email{shuaiw@cse.ust.hk}

\author{Yuan Luo}
\affiliation{%
  \institution{Shanghai Jiao Tong University}
  \city{Shanghai}
  \country{China}
}
\email{yuanluo@sjtu.edu.cn}

\author{Pingchuan Ma}
\affiliation{%
  \institution{Zhejiang University of Technology}
  \city{Hangzhou}
  \country{China}
}
\email{pma@zjut.edu.cn}

\author{Yuxuan Du}
\affiliation{%
  \department{College of Computing and Data Science, and
    School of Physical and Mathematical Sciences}
  \institution{Nanyang Technological University}
  \city{Singapore}
  \country{Singapore}
}
\email{yuxuan.du@ntu.edu.sg}

\authornote{Corresponding authors: Chenhao Ying, Yuan Luo, Pingchuan Ma and Yuxuan Du.}

\authornote{Both authors contributed equally to this research. \\
This work was supported by Grant RS3/26 from the Ministry of Education (MOE), Singapore; the RGC GRF grant under the contract 16214723; and the National Natural Science Foundation of China (NSFC) under Grant 62402313.
}

\renewcommand{\shortauthors}{Xia et al.}

\makeatletter

\def\@authornotemark{}
\def\@@authornotemark#1{}

\def\@mkauthors{%
  \begingroup
  \hsize=\textwidth
  \global\setbox\mktitle@bx=\vbox{%
    \unvbox\mktitle@bx

    \centering

    {\normalsize
      \resizebox{0.98\textwidth}{!}{%
        \begin{tabular}{@{}c@{}}
          Fuyuan Xia\textsuperscript{1,\textdagger}
          \quad
          Qixin Zhang\textsuperscript{2,\textdagger}
          \quad
          Chenhao Ying\textsuperscript{1,*}
          \quad
          Haojin Zhu\textsuperscript{1}
          \quad
          Shuai Wang\textsuperscript{3}
          \quad
          Yuan Luo\textsuperscript{1,*}
          \quad
          Pingchuan Ma\textsuperscript{4,*}
          \quad
          Yuxuan Du\textsuperscript{2,*}
        \end{tabular}%
      }\par
    }%

    {\small\itshape
      \begin{tabular}{@{}c@{}}
        \textsuperscript{1}Shanghai Jiao Tong University,
        \{fuyuanxia, yingchenhao, zhu-hj, yuanluo\}@sjtu.edu.cn
        \\

        \textsuperscript{2}Nanyang Technological University,
        \{qixin.zhang, yuxuan.du\}@ntu.edu.sg
        \\

        \textsuperscript{3}Hong Kong University of Science and Technology,
        shuaiw@cse.ust.hk
        \\

        \textsuperscript{4}Zhejiang University of Technology,
        pma@zjut.edu.cn

      \end{tabular}\par
    }%

    \vspace{0.25em}%
  }%
  \endgroup
}

\makeatother

\renewcommand{\shortauthors}{Xia et al.}

\setlength{\textfloatsep}{2pt plus 1pt minus 1pt}
\setlength{\dbltextfloatsep}{2pt plus 1pt minus 1pt}
\setlength{\floatsep}{2pt plus 1pt minus 1pt}
\setlength{\intextsep}{2pt plus 1pt minus 1pt}
\setlength{\abovecaptionskip}{2pt plus 0.5pt minus 0.5pt}
\setlength{\belowcaptionskip}{2pt plus 0.5pt minus 0.5pt}
\setlength{\abovedisplayskip}{2pt plus 0.5pt minus 0.5pt}
\setlength{\belowdisplayskip}{2pt plus 0.5pt minus 0.5pt}

\begin{abstract}
As quantum computing continually improves, ensuring the reliability and correctness of quantum libraries has become increasingly critical. To this end, many LLM-based fuzzing approaches towards quantum libraries have been proposed to uncover potential bugs. However, these methods still suffer from limitations such as insufficient flexibility and low efficiency, which hinder the progress of the quantum computing field.
To address these challenges, we propose \textbf{\tool}, a novel knowledge-guided fuzzer for quantum libraries. It leverages comprehensive codebase knowledge to ground LLM-based test generation, synergizing this with fitness-guided evaluation and two-level mutations to explore complex execution paths and trigger potential bugs. Firstly, \tool introduces a novel prompting scheme tailored to quantum programs, which strategically incorporates knowledge of the codebase to efficiently generate high-quality quantum seed programs. Moreover, we develop evaluation and mutation strategies to handle the generated seed programs, facilitating efficient fuzzing execution while further enriching the diversity of the resulting test cases.
We implement \tool and conduct fuzzing on three popular quantum libraries, including Qiskit, PennyLane, and Cirq. Experimental results demonstrate that our approach significantly outperforms other state-of-the-art methods, with coverage improved by up to 18.44\%. During the development of \tool, we discovered 13 bugs, all of which have been confirmed and 12 have already been fixed by the developers.
\end{abstract}


\maketitle

\section{Introduction}\label{sec:intro}

Diverse quantum libraries have been developed by academic institutions and technology companies to accelerate the design and implementation of quantum algorithms, motivated by the promise that quantum computing offers capabilities beyond classical computation across diverse domains~\cite{quantum,ying2025tosem}. However, an often overlooked issue is that bugs in such quantum libraries may produce unexpected results, leading to flawed scientific or engineering conclusions and obscuring the true potential of quantum computing~\cite{ali2022taoyue,fang2024pldi,assolini2024sas,li2026methodological,long2024tosem,upadhyay2026understandingbugsquantumsimulators}. Scarce quantum hardware further amplifies the impact of software bugs, leading to significant resource waste. In this regard, much like in classical software engineering, there is an increasing need for systematic quantum library testing to ensure correctness and reliability.

However, porting classical testing methodologies to the quantum domain is far from straightforward. Conventional techniques like static analysis~\cite{staticanalysis} often fail because they lack the domain-specific semantics required to comprehend complex quantum states, making verification difficult. Furthermore, the rapid evolution of quantum libraries quickly renders manual testing rules obsolete, necessitating automated and adaptive approaches that can systematically explore the complex logic of these systems. To meet these requirements, fuzzing~\cite{manes2019tse,jiang2024fuzzing} has emerged as a leading candidate for automatically discovering bugs. By generating and executing a vast number of randomized test cases, fuzzing can effectively explore the complex input space and identify unexpected software behaviors across diverse execution scenarios, which is uniquely suited for quantum libraries.

\begin{figure}
\centering
\includegraphics[width=0.94\linewidth]{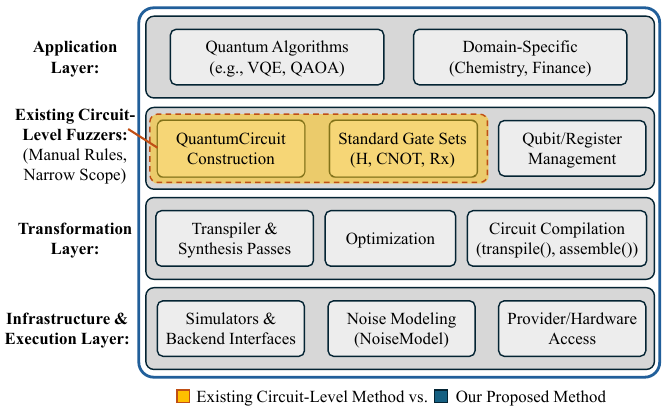}
\caption{Comparison of Fuzzing Scopes.} 
\label{fig:intro}
\end{figure}

\parh{Limitations of existing fuzzing-based quantum library testing.} In this work, quantum-library testing treats the library or platform implementation as the system under test, while quantum programs serve as test inputs. Existing quantum-library fuzzers mainly rely on domain-specific test generation~\cite{paltenghi2023morphq,shaking2025oopsla,hu2024issta}, using dedicated generators, transformation rules, templates, or constraint specifications to construct structurally valid programs. However, these approaches necessitate extensive domain expertise to manually formulate generation and mutation rules. Moreover, as shown in Figure~\ref{fig:intro}, such manual curation confines their scope to circuit-related APIs, rendering them inflexible and unable to exercise emerging software features or high-level orchestration logic. Consequently, they are incapable of identifying entire classes of bugs in the rapidly evolving quantum computing ecosystem.

Alternatively, general-purpose LLM-based fuzzers such as~\cite{xia2024fuzz4all} offer a promising path forward, as they can use their domain knowledge to generate seeds invoking diverse APIs without manual rule design. However, instantiating LLM-based fuzzing for quantum programming libraries faces unique \textbf{C}hallenges compared to classical software.\\
\textbf{C1. Low-validity seed generation.} Unlike classical software with stable interfaces, the frequent restructuring of quantum APIs hinders LLMs from maintaining up-to-date internal representations of the evolving API surface and its associated semantic constraints. As illustrated in Figure~\ref{fig:fuzzing_eval} in \S~\ref{sec:challenge}, our pilot study reveals that LLMs achieve only 35--46\% validity for quantum programming libraries compared to 64--86\% for classical programming libraries with disproportionately high import and attribute errors, averaging 118 such errors per 600 programs in each library.\\
\textbf{C2. Constrained exploration in quantum programs.} Unlike classical software, quantum programs are subject not only to hardware-level constraints such as limited qubit counts and restricted connectivity, but also to framework-level execution semantics and hybrid quantum–classical control flows. Conventional mutation strategies operating at the byte or abstract-syntax-tree level often fail to preserve these structural and semantic constraints, thereby generating invalid or non-executable programs. The core challenge is therefore to design domain-aware mutation mechanisms that systematically explore the input space while respecting quantum-specific structural, semantic, and complicated constraints.

\parh{Our solution.}~To address these challenges, we propose \textbf{\tool}, a \underline{K}nowledge-guided \underline{Q}uantum \underline{Fuzz}ing framework, which incorporates codebase-specific structural and semantic information to guide LLM-driven seed program generation for quantum libraries. Moreover, \tool integrates fitness-guided evaluation and mutation mechanisms to systematically explore complex execution paths and expose latent defects. 

To address \textbf{C1}, \tool constructs an
API corpus capturing four types of knowledge: (1) static metadata, such as signatures and
locations; (2) API associations via three dimensions, \emph{i.e.}, proximity, type coupling,
and call relationships; (3) semantic models via LLM-based summarization; and (4)
evolution metrics tracking API modifications across versions. During generation,
\tool employs a probabilistic selection strategy that prioritizes APIs with strong semantic associations and high evolutionary activity, steering the LLM to synthesize valid, version-aligned \emph{seed programs} that target potentially bug-prone execution paths. 

In response to \textbf{C2}, \tool employs a two-level mutation strategy, where parameter-level mutation targets numerical corner cases and gate-level structural mutation explores variations through entanglement-aware substitutions.
Furthermore, we introduce a fitness function that integrates gate diversity, entangled qubits, API diversity, and call depth. This function acts as a selection criterion to prioritize high-quality seeds, steering the LLM toward generating programs in more complex and potentially bug-prone regions.
Seeking a balance between precision and throughput, \tool adopts a tiered synergy: a high-capacity model (\emph{i.e.}, Gemini-3~\cite{team2023gemini}) performs one-time API semantic modeling to ground a cost-effective LLM (\emph{e.g.}, Qwen 2.5-Coder) for iterative fuzzing generation. Finally, we execute the resulting \emph{test cases}—comprising both the generated seed programs and their mutated variants—and apply heuristic filtering to the execution results to identify impactful bugs within the target quantum libraries.

We evaluate the performance of \tool on three mainstream quantum libraries:
Qiskit\footnote{\url{https://github.com/Qiskit/qiskit}},
PennyLane\footnote{\url{https://github.com/PennyLaneAI/pennylane}}, and
Cirq\footnote{\url{https://github.com/quantumlib/Cirq}}. \tool significantly
outperforms existing baselines in code validity and bug-finding effectiveness,
discovering 13 bugs, all of which have been confirmed and 12 already fixed by the developers. Our contributions are summarized as follows.
\begin{itemize}
    \item We propose \tool, a novel knowledge-guided quantum fuzzing framework that integrates codebase-specific insights with LLM-based test generation and domain-specific mutation to systematically explore complex execution paths in quantum libraries.
    \item We construct a comprehensive API corpus encompassing four dimensions of knowledge. Based on this, a seed program generation strategy is designed to steer the LLM toward generating valid, version-aligned, and potentially bug-prone quantum programs.
    \item We introduce a two-level mutation strategy (parameter and gate-level) and a multi-dimensional fitness function that accounts for gate diversity and entanglement. This approach prioritizes high-quality seeds to effectively navigate the complex state space of quantum programs.
    \item We perform a comprehensive evaluation of \tool on three mainstream quantum libraries. The results demonstrate its superiority over existing baselines, surpassing the state-of-the-art coverage by a margin of up to 18.44\%, successfully identifying 13 new bugs, all of which have been confirmed by developers and 12 of which are already fixed.
\end{itemize}

\section{Background}

\subsection{Basic Concepts in Quantum Computing}

Quantum computing leverages principles such as superposition and entanglement to
enable new computational capabilities~\cite{gill2025quantumc}. In practice, these capabilities are commonly formulated within the \emph{quantum circuit model}~\cite{fowler2012surface}, in which computation is represented as a sequence of elementary quantum gates, \emph{e.g.}, single-qubit and entangling two-qubit operations, acting on quantum states. After that, a measurement process is applied to extract information from the final quantum state into classical registers, yielding probabilistic outcomes that encode the solution to a given problem~\cite{abbas2024nrp,cerezo2021nrp}. Refer to Supplementary Material (\textbf{SM})~A for more details.

Within the quantum computing ecosystem, Qiskit~\cite{qiskit}, PennyLane~\cite{pennylane}, and Cirq~\cite{cirq_developers_2025_10.5281/zenodo.4062499} are three major quantum programming libraries maintained by major technology companies or quantum-focused companies. 
While these libraries share extensive functional commonalities in circuit synthesis, simulation, and hardware integration, they provide alternative development environments that cater to different hardware-specific optimizations and research workflows.
Despite their widespread adoption, these libraries remain in a state of perpetual and volatile transformation to keep pace with rapid hardware and algorithmic breakthroughs~\cite{paltenghi2023morphq}. These rapid shifts often result in hidden defects, leading to persistent logic anomalies and unexpected behaviors that deviate from intended specifications. These recurring bugs represent a significant bottleneck that necessitates rigorous testing and validation to ensure the integrity of the quantum computing ecosystem.

\subsection{Fuzzing}

\parh{Fuzzing foundations and LLM-based evolution.} Fuzzing is a cornerstone of automated software testing, designed to identify bugs and edge-case behaviors by executing a target system with a vast array of synthesized inputs~\cite{fioraldi2020afl,stephens2016driller,bohme2017directed,zhu2022csur,she2024ccs,xie2025ase,bohme2020boosting,wu2022one}. Traditional fuzzing taxonomies primarily distinguish between generation-based approaches, which construct inputs from predefined grammars, and mutation-based strategies, which apply stochastic transformations to existing seeds to maximize code coverage~\cite{dong2025tdsc,manes2019tse}. While effective for low-level protocols, these conventional methods often struggle with high-level software that requires complex, semantically valid structures. Recently, LLMs have redefined this paradigm. By leveraging their inherent semantic comprehension and code synthesis capabilities, LLM-driven fuzzers can autonomously generate syntactically sophisticated and contextually rich test cases~\cite{xia2024fuzz4all,deng2024icse}. This shift alleviates the burden of manual rule engineering and enables the exploration of deep logic within complex API sequences.

\parh{Fuzzing in the quantum libraries.} The application of fuzzing to quantum libraries has bifurcated into two primary paradigms: circuit-level construction and LLM-driven exploration. Circuit-level fuzzers typically employ domain-specific generators, transformation rules, or templates to generate gate sequences, ensuring high validity but often remaining confined to low-level gate logic~\cite{paltenghi2023morphq,shaking2025oopsla}. In contrast, LLM-based methods harness extensive semantic knowledge to probe high-level behaviors in quantum programming libraries. A persistent objective in this field is to effectively bridge the gap between the expressive flexibility of LLM-generated test cases and the stringent syntactic requirements of quantum libraries, ensuring that the generated programs remain both diverse and executable.

\section{Motivation \& Challenges}
\begin{figure}[t]
    \centering
    \begin{minipage}{8.4cm}
        \begin{lstlisting}
# <IMPORT TEMPLATE>
qc = QuantumCircuit(11, 11, name='qc')
qc.append(ECRGate(), qargs=[qc.qubits[2], qc.qubits[5]], cargs=[])
qc.append(CHGate(), qargs=[qc.qubits[5], qc.qubits[7]], cargs=[])
qc.append(ZGate(), qargs=[qc.qubits[5]], cargs=[])
qc.append(ZGate(), qargs=[qr[7]], cargs=[])
qc.append(RCCXGate(), qargs=[qr[8], qr[2], qr[6]], cargs=[])
qc.append(ZGate(), qargs=[qr[0]], cargs=[])
qc.append(iSwapGate(), qargs=[qr[1], qr[3]], cargs=[])
qc.append(SdgGate(), qargs=[qr[2]], cargs=[])
qc.append(CXGate(), qargs=[qr[3], qr[1]], cargs=[])
# <RESULT TEMPLATE>        \end{lstlisting}

        \subcaption{Code Generated by Circuit-Level Method}
    \end{minipage}

    \begin{minipage}{8.4cm}
        \begin{lstlisting}
# <QUANTUM CODE>
from qiskit.circuit.library import grover_operator
oracle = QuantumCircuit(3)
oracle.h(2)
oracle.ccx(0, 1, 2)
oracle.h(2)
grover_op = grover_operator(oracle) # Involved API
qc = QuantumCircuit(3)
qc.h([0, 1, 2])
qc.append(grover_op, [0, 1, 2])
qc.measure_all()
# <QUANTUM CODE>       \end{lstlisting}
        \subcaption{Realistic Quantum Program}
    \end{minipage}

    \caption{Comparison between circuit-level fuzzer-generated code and realistic quantum program.}\label{fig:example_diff_method}
\end{figure}

\subsection{Revisiting Fuzzing for Quantum Libraries}
We summarize two typical families of existing solutions for fuzzing quantum libraries as follows. First, \textbf{circuit-level fuzzers} aim to generate random quantum circuits to trigger bugs in the quantum software stack~\cite{paltenghi2023morphq,shaking2025oopsla}. Their primary approaches use domain-specific generators, transformation rules, templates, or constraint specifications to rapidly produce valid quantum programs, prioritizing the efficient discovery of structural defects such as crashes or incorrect circuit outputs. Nevertheless, this design entails an inherent trade-off between efficiency and generality. By enforcing predefined structural rules to ensure efficient and valid circuit generation, such approaches limit their applicability to a broader range of quantum programs. Consequently, these approaches face two key challenges: (1) manually designed rules become increasingly difficult to maintain as quantum programming libraries evolve and introduce new operations; and (2) their highly specialized generation mechanisms primarily target structural circuit anomalies, limiting their ability to expose the diverse defects that arise in complex and real-world quantum libraries. As indicated in Figure~\ref{fig:example_diff_method}(a), MorphQ~\cite{paltenghi2023morphq} can generate legal circuits containing a large number of random quantum gates. However, in practice (\emph{e.g.}, the deployment of the Grover algorithm on real quantum hardware), additional APIs beyond circuit gates are required, such as \texttt{grover\_operator} in Figure~\ref{fig:example_diff_method}(b). For such APIs that go beyond the circuit level, potential bugs cannot be uncovered by these methods, thus constraining the applicability of circuit-level fuzzers to quantum libraries.

While fuzzers at the circuit level often suffer from limited flexibility, approaches based on LLMs have emerged as a promising alternative~\cite{deng2023issta,xia2024fuzz4all}. Attributed to the extensive pretraining on diverse code corpora, LLMs have the potential to effectively capture the complex semantic logic of quantum libraries to generate diverse test cases. However, \textit{the syntactic and semantic validity of quantum code generated by LLMs remains a significant bottleneck}. As reported in~\cite{xia2024fuzz4all}, only 24.9\% of Qiskit code generated by LLMs is valid, a figure that pales in comparison to the 100\% validity rate achieved by circuit-level fuzzers such as MorphQ~\cite{paltenghi2023morphq}. This high failure rate severely constrains the practical utility of LLMs in the domain of quantum library testing.

\subsection{Motivating Study}
\label{sec:challenge}
As discussed in (\S~\ref{sec:intro}), LLMs can be leveraged to generate fuzzing inputs targeting a given library. To systematically assess the capability of LLMs to generate fuzzing inputs for quantum libraries, we conduct a preliminary empirical comparison between quantum libraries and their well-established classical analogues.

\parh{Experimental configuration.}
We select three prominent quantum libraries, Qiskit, Cirq, and PennyLane, as our primary targets. To provide a comparative baseline, we also include three widely used classical libraries: Numpy\footnote{\url{https://github.com/numpy/numpy}}, Pandas\footnote{\url{https://github.com/pandas-dev/pandas}}, and PyTorch\footnote{\url{https://github.com/pytorch/pytorch}}. We adopt a representative LLM-based fuzzing pipeline similar to~\cite{lin2025ase}, utilizing a Chain-of-Thought (CoT)~\cite{cot} prompting paradigm to guide the models through library importation, data preparation, and API invocation. For each library, we randomly sample 300 distinct APIs and employ two open-source models, CodeLlama-13B~\cite{codellama} and Qwen2.5-Coder-14B~\cite{qwen2.5-coder}, to generate one test case per API. We select models at this scale as they strike an optimal balance between robust reasoning capabilities and local reproducibility. This setup yields a total of 600 test cases per library, allowing us to evaluate the models' zero-shot generation capabilities without domain-specific knowledge.

\begin{figure}[t]
  \centering
  \includegraphics[width=0.96\linewidth]{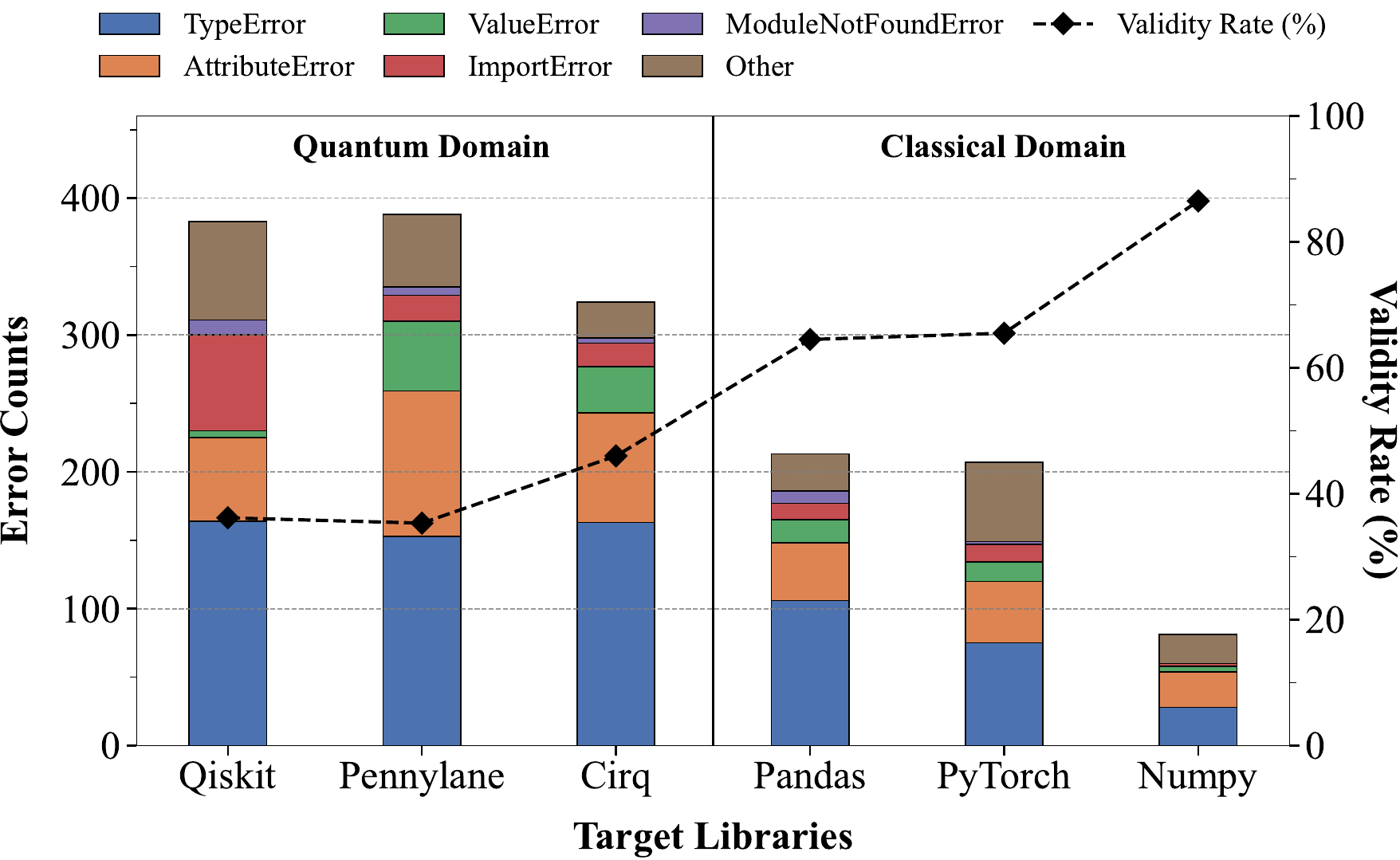}
  \caption{Empirical evaluation of the validity and error patterns in LLM-generated fuzzing test cases across quantum and classical libraries. The bars represent absolute error counts (left y-axis), while the dashed line indicates the validity rate (right y-axis).}
  \label{fig:fuzzing_eval}
\end{figure}

\parh{Quantifying the performance gap.}
As illustrated in Figure~\ref{fig:fuzzing_eval}, our results reveal a stark disparity in code validity between domains. While classical libraries achieve a pass rate of 64.50\% to 86.50\%, the validity for quantum libraries plummets to a range of 35.33\% to 46.00\%. Our empirical results in Figure~\ref{fig:fuzzing_eval} show that \textit{TypeError}, \textit{AttributeError}, and \textit{ValueError} are the most frequent errors (\emph{e.g.}, 164 TypeErrors for Qiskit), while \textit{ImportError} and \textit{ModuleNotFoundError} are also common. These findings demonstrate that despite their strong general coding capabilities, LLMs exhibit a pronounced lack of domain awareness when encountering quantum-specific syntax and semantics.

\begin{figure}
\centering
\includegraphics[width=3.2in]{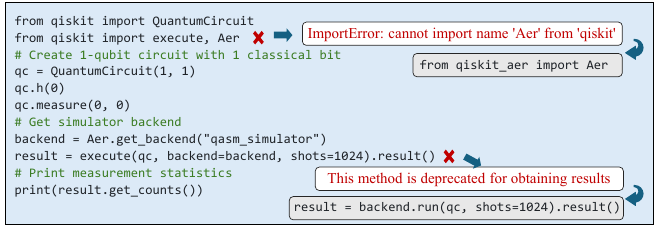}
\caption{Code example of typical errors in quantum computing libraries generated by LLMs.}\label{fig:crash_show}
\end{figure}

\parh{Root cause analysis.}
A representative example illustrated in Figure~\ref{fig:crash_show} reveals that such failures are primarily attributed to the volatile evolution and frequent breaking changes inherent in quantum libraries. 
Unlike classical domains where software updates are primarily driven by application logic, these reconfigurations are uniquely induced by \textit{the rapid advancement of quantum hardware}.
As new functionalities and physical architectures emerge, library maintainers must frequently reconfigure invocation patterns and structural placements to accommodate hardware-side capabilities. Such instability leads to inconsistent internal representations within LLMs, heightening their proclivity for generating deprecated or hallucinated code. Furthermore, unlike the mature classical computing ecosystem, there is a pronounced scarcity of publicly available codebases aligned with the latest versions of quantum libraries. Consequently, the available resources for knowledge acquisition are largely restricted to the libraries themselves and the LLM’s own (often flawed) outputs. These observations motivate us to investigate the following guiding questions: \ding{182} \textit{How can library-specific knowledge be leveraged to guide LLMs in generating valid seed programs,} and
\ding{183} \textit{how can these seeds be systematically evolved to expose latent bugs in quantum libraries?}

\subsection{Challenge}
To address the aforementioned guiding questions, a fundamental prerequisite is to understand the key challenges that arise when applying LLM-based fuzzing to quantum libraries. These challenges are inherently progressive: generating valid, library-aligned test cases is the foundation, upon which effective exploration through mutation must be built.

\begin{figure*}
\centering
\includegraphics[width=0.99\linewidth]{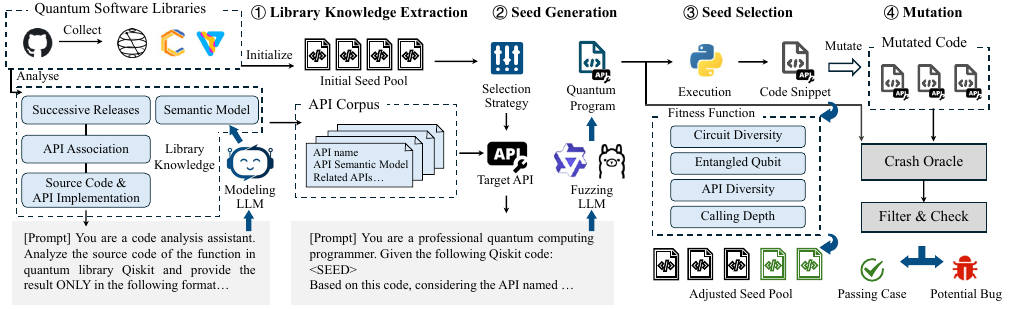}
\caption{The workflow of \tool for detecting bugs in quantum libraries.}\label{fig:workflow}
\end{figure*}

\parh{Challenges in generating valid quantum seeds.} 
LLMs struggle to generate valid test cases for quantum libraries due to the rapid evolution and frequent restructuring of APIs, which lead to outdated or hallucinated invocations. This issue is further exacerbated by the scarcity of up-to-date quantum code, limiting the model’s ability to learn stable usage patterns and diverse invocation semantics. Furthermore, since fuzzing is a resource-intensive process requiring high-volume iterations, a fundamental challenge lies in balancing computational cost with testing effectiveness. As a result, seed generation becomes a knowledge-grounding problem, requiring alignment with current library implementations (\emph{i.e.}, source code), while also balancing cost and effectiveness for large-scale fuzzing with open-source models.

\parh{Constraints and inefficiencies in quantum exploration.} 
The subsequent exploration phase remains challenging even with valid seeds. Effective exploration of the program space typically relies on an iterative process of mutation, fitness evaluation, and input regeneration. However, directly applying traditional fuzzing strategies in the quantum setting results in severely limited exploration. First, at the mutation level, quantum programs are governed by hardware-imposed structural constraints and framework-defined execution semantics. Generic mutation operators (\textit{e.g.}, byte-level or AST-level transformations) often fail to preserve these constraints, producing invalid programs and significantly reducing the proportion of executable test cases. Second, at the evaluation and selection level, even when syntactically and structurally valid variants are generated, the lack of domain-specific fitness metrics limits the ability to effectively prioritize high-value seeds. Without quantum-aware fitness functions to measure traits like quantum circuit complexity, the fuzzer cannot prioritize high-potential candidates. Consequently, the entire exploration loop becomes inefficient, trapped in generating redundant inputs with limited capability to evolve simple seeds into complex test cases. In this regard, holistic and domain-specific adaptations across the mutation and selection mechanisms are highly demanded to expose latent bugs.

\section{Approach}
\label{sec:approach}
KQFuzz addresses a central challenge in quantum-library fuzzing: generating tests that remain executable under evolving APIs while being diverse enough to exercise deep library logic. Existing circuit-level fuzzers provide high validity but limited API coverage, whereas general-purpose LLM fuzzers cover broader APIs but often rely on outdated API knowledge and generate invalid tests. Moreover, generic mutation and selection strategies lack quantum-aware guidance for parameterized gates, entanglement structures, and multi-API interactions.

To bridge this gap, KQFuzz employs a four-stage validity-to-diversity pipeline as illustrated in \F~\ref{fig:workflow}. First, in \textbf{Library Knowledge Extraction} (Stage \textcircled{1}, \S~\ref{sec:know_extra}), \tool constructs an API corpus by extracting structural associations, semantic profiles, and evolution metrics from the library code. Subsequently, during the \textbf{Seed Generation} (Stage \textcircled{2}, \S~\ref{sec:seed_gen}) phase, a Fuzzing LLM takes seeds from the pool and incrementally incorporates target APIs to generate diverse quantum programs. Next, in \textbf{Seed Selection} (Stage \textcircled{3}, \S~\ref{sec:seed_selection}), the generated programs are executed and assessed via a multi-dimensional fitness function. The generated quantum programs with high semantic complexity are fed back into the adjusted seed pool for the next iteration. Finally, the \textbf{Mutation} (Stage \textcircled{4}, \S~\ref{sec:mutation_oracle}) phase systematically transforms quantum programs to expand the search space, while a crash oracle and semantic filters are employed to identify potential bugs.

\subsection{Knowledge Modeling from Library Code}
\label{sec:know_extra}

As identified in our pilot study (\S~\ref{sec:challenge}), LLM-based fuzzers
frequently generate invalid seed programs due to outdated library knowledge or
hallucinated API usage. To mitigate these issues, \tool constructs an explicit
knowledge model directly from the source code of the target quantum library.
Rather than relying on parametric knowledge embedded in the LLM, \tool extracts
lightweight, library-specific structural and semantic cues and organizes them
into an API corpus that guides subsequent seed generation.

\parh{API inventory and static metadata.} 
\tool begins by constructing an inventory of available APIs from the official
documentation of the target software. Let $\mathcal{I} = \{1, 2, \ldots, N\}$
denote the resulting set of APIs. For each API $i \in \mathcal{I}$, \tool
collects static metadata by inspecting the corresponding source files, including
the API signature, its implementation body, and its file and module location.
This information serves as the foundation for modeling relationships among APIs
and for providing accurate and version-aligned context to the LLM during seed
synthesis.

\parh{API association modeling.}
To capture semantic and structural relationships between APIs, \tool models pairwise API associations using three complementary dimensions. These dimensions are designed as lightweight signals derived from source-level inspection. In particular, \textit{Proximity}, denoted by $S_p(a,b)$, measures the structural locality via file and module boundaries; \textit{Type Overlap}, denoted by $S_t(a,b)$, quantifies the shared domain-specific input and output types; and \textit{Call Relationships}, denoted by $S_c(a,b)$, evaluates direct reachability and shared invocation patterns. We aggregate these three dimensions into a joint association score, \textit{i.e.},
\begin{equation}
\label{eq:score_ab}
S(a,b) = w_1 S_p(a,b) + w_2 S_t(a,b) + w_3 S_c(a,b),
\end{equation}
where $w_1$, $w_2$, and $w_3$ control the relative importance of these dimensions. Detailed mathematical formulations for each dimension are provided in SM~B.

\parh{API implementation of semantic modeling.} \tool introduces an API implementation of a semantic model by prompting an LLM with source code to infer descriptions, constraints, outputs, and fuzzing points (detailed prompt in SM~C). The modeling LLM is limited to this semantic summarization; API inventory construction and the computation of association and evolution metrics remain non-LLM source-analysis tasks. The resulting semantic model is a structured semantic summary grounded in the library source code. Unlike resource-intensive seed generation, semantic extraction is a \textit{one-time preprocessing step}. We employ Gemini-3~\cite{team2023gemini} to synthesize library profiles, using its extended context window to capture long-range dependencies in complex codebases. To clarify the distinct roles of LLMs within our framework, we designate the LLM used for this extraction as the \emph{modeling LLM}, while the LLM responsible for subsequent seed generation is referred to as the \emph{fuzzing LLM}.

\parh{API evolution modeling.} In addition to structural associations, \tool
models API evolution across library versions. APIs that undergo frequent
modifications may expose unstable or under-tested behaviors and thus represent
promising fuzzing targets. For a given API $i\in \mathcal{I}$, denote $B_i^\prime$ and $B_i$
as the implementation in two successive versions, with $B_i^\prime$
preceding $B_i$. We quantify the evolution weight $M'(i)$ as
\[
M'(i) = 1 - \frac{|\mathcal{G}_n(B_i) \cap \mathcal{G}_n(B_i^\prime)|}{|\mathcal{G}_n(B_i) \cup \mathcal{G}_n(B_i^\prime)|},
\]
where $\mathcal{G}_n(\cdot)$ extracts $n$-gram sets from the \emph{normalized Abstract Syntax Tree (AST) node sequences} of the implementations. To further prioritize unstable APIs, we map $M'(i)$ to the degree of modification:
\begin{equation}\label{eq:Mi}
    M(i) = 1 + \delta \cdot \tanh(\gamma(M'(i) - 0.5)).
\end{equation}
This non-linear mapping amplifies the testing priority of high-evolution APIs while suppressing that of stable ones, where $\gamma$ and $\delta$ are hyper-parameters controlling the sensitivity and influence range, respectively. 

\noindent\textit{Remark}. All extracted metadata, association scores, semantic model, and evolution metrics are integrated into a unified API corpus, which is subsequently used to guide API selection and prompt construction during seed generation for fuzzing.

\subsection{LLM-Guided Seed Generation}
\label{sec:seed_gen}
Building on the API corpus constructed in (\S~\ref{sec:know_extra}), \tool
iteratively generates seed programs for fuzzing by combining probabilistic API
selection with LLM-based code synthesis. The key idea is to incrementally extend
existing quantum programs with carefully selected APIs, while using
library-specific knowledge to constrain and guide the LLM toward valid and
semantically meaningful code.

\parh{Probabilistic target API selection.}
Given a seed program containing an invocation of API $i\in \mathcal{I}$, \tool queries the API
corpus to retrieve the set of related APIs, \textit{i.e.},
\[
\mathcal{S}_i = \{ j \in \mathcal{I} \mid S(i,j) \neq 0 \},
\]
where $S(i,j)$ is the association score defined in \E~(\ref{eq:score_ab}).
Instead of selecting a related API uniformly at random, \tool leverages the
metrics obtained from knowledge modeling to bias selection toward APIs that are
both strongly associated with the seed API and actively evolving. Specifically,
it assigns each candidate API $k \in \mathcal{S}_i$ a selection probability:
\begin{equation}
\label{eq:select_pr}
\text{Pr}[k;i] =
\frac{M(k) \cdot S(i,k)}
{\sum\nolimits_{j \in \mathcal{S}_i} M(j) \cdot S(i,j)} ,
\end{equation}
where $S(i, k)$ and $M(k)$ are defined in Eqn.~\ref{eq:score_ab} and Eqn.~\ref{eq:Mi}, respectively. As a result, \tool prioritizes
semantically meaningful API combinations while avoiding the forced integration
of unrelated APIs that often lead to invalid or trivial programs.

\parh{Prompt construction and code generation.} Given a selected seed program
and a target API $i_t\in\mathcal{I}$ sampled according to the probabilities in
\E~(\ref{eq:select_pr}), \tool invokes the fuzzing LLM to synthesize a new seed program
that incorporates $i_t$ into the seed program. The goal of prompt construction
is to (i) ground the LLM in the \emph{exact} library version under test, (ii)
provide sufficient semantic context to avoid hallucinated usage, and (iii) steer
generation toward either introducing $i_t$ or exercising it thoroughly, depending on the stage of fuzzing.

Accordingly, \tool constructs prompts by integrating three key sources of information: (i) the selected seed program, which provides the execution context and structural skeleton; (ii) the semantic correlation between existing APIs in the seed and the target API $i_t$, ensuring that the newly introduced API is contextually relevant; and (iii) the API implementation semantic model of $i_t$ (detailed prompt in SM~C). Unlike approaches that rely on the LLM's internal parametric knowledge, \tool provides an explicit knowledge representation synthesized from the source code. This semantic model encapsulates inferred descriptions, argument constraints, expected outputs, and critical fuzzing points. By presenting these structured semantic profiles, \tool explicitly grounds the generation process in the exact library logic, exposing intricate dependencies and usage patterns that are often absent from outdated training data.
Moreover, by promoting the generation of programs that integrate multiple semantically related APIs, \tool produces seed programs that exercise more complex and diverse execution behaviors, thereby increasing coverage of the target quantum library.

\parh{Comparison with standard RAG.} While the integration of an API corpus shares the high-level philosophy of Retrieval-Augmented Generation (RAG)~\cite{lewis2020retrieval}, namely augmenting LLMs with external knowledge, our mechanism differs fundamentally from standard RAG pipelines. Specifically, standard RAG relies on vectorizing documents and retrieving text chunks based on semantic similarity to a user query. However, in the highly constrained domain of quantum libraries, semantic similarity alone does not ensure structural compatibility or valid program execution. In addition, instead of treating knowledge as isolated text snippets, our approach constructs a structured and multi-dimensional knowledge model that captures API relationships, semantic properties, and historical evolution. This design enables the fuzzer to preserve structural constraints while prioritizing components with higher defect potential. As a result, our approach provides a more principled and robust basis than conventional text-based retrieval methods.

\subsection{Seed Selection}
\label{sec:seed_selection}

\tool follows an iterative generation-selection loop. At each iteration, a batch
of new seed programs is generated from the current seed pool, after which
high-quality programs are selected and promoted as seeds for subsequent
iterations. Below, we separately describe how the seed pool is initialized and how the generated programs are prioritized.

\parh{Seed pool initialization.}
\tool initializes the seed pool using officially provided usage examples
extracted from API docstrings of the target quantum library. The number of
initial seeds is comparable to the number of APIs, ensuring broad coverage of
library functionality. Importantly, all seed programs are derived exclusively
from the target library itself, avoiding reliance on external quantum code that
may be outdated or incompatible.

\parh{Fitness-guided seed selection.}
\tool evaluates generated seed programs using a
fitness function and prioritizes high-quality programs for reseeding. For each
generated program, \tool first extracts the largest executable fragment using a
greedy strategy, ensuring that fitness is computed only on runnable code. For example, consider a 15-line program that crashes at line 12. In this case, the fragment consisting of the first 11 lines constitutes the largest executable fragment. It then analyzes the structure of each generated seed program and assigns a fitness score that quantifies its exploration potential. The fitness function is
designed to favor programs that exercise complex quantum operations and diverse
API interactions. Specifically, \tool evaluates each program along the following
dimensions: \textit{\textbf{Circuit Gate Diversity ($G$).}} The number of
distinct quantum gate types appearing in the circuit. Higher diversity indicates
broader coverage of quantum operations, increasing the likelihood of exposing
bugs triggered by specific gate combinations. \textit{\textbf{Number of
Entangled Qubits ($Q$).}} The number of qubits participating in entangled
operations. Entanglement amplifies circuit complexity and may expose subtle
compiler or execution errors, although excessively large entangled states may
incur the hardness of efficient simulation. \textit{\textbf{API Diversity ($\Delta$).}} The number of
distinct APIs invoked in the quantum program. This metric reflects semantic coverage
across different library components, such as circuit construction, simulation,
and backend configuration. \textit{\textbf{API Calling Depth ($L$).}} The
maximum depth of API call chains, capturing the temporal and logical complexity
of API interactions. Longer call chains stress long execution paths and stateful
behaviors. Combining these factors, the fitness score of a generated program $C$
is defined as
\begin{equation}
    \text{FitnessScore}(C) = G + \min(Q,\tau) + L + \eta\cdot\Delta .
\end{equation}
To prevent the fuzzer from over-optimizing for entanglement at the expense of executability, the term $Q$ is capped at $\tau$. This design penalizes excessively deep entanglement that threatens to crash the target library. Furthermore, $\eta$ ($\eta \ge 1$) serves as a tunable weight parameter designed to balance the diversity of API invocations ($\Delta$) against the other evaluation metrics, ensuring that API diversity is appropriately prioritized without overwhelming the fitness score. 

In each iteration, \tool selects the top-$\kappa$ programs and adds them to the seed
pool for the next round. This fitness-guided prioritization biases
the fuzzing process toward semantically rich test
cases, enabling progressively broader and more complex exploration of the target quantum library over time.

\subsection{Mutation and Test Oracle}
\label{sec:mutation_oracle}
Following LLM-guided generation, \tool expands the search space through structured mutation and utilizes a crash-based test oracle for bug detection. Mutation
complements generation by systematically exploring semantic variations that are
unlikely to be produced by prompting alone, while the test oracle provides
automated bug detection indicators during execution.

\parh{Mutation.} \tool applies semantics-aware mutations to generated test cases
by directly operating on quantum circuits. The mutation strategy consists of two
levels: parameter-level mutation and gate-level structural mutation.

\noindent\textit{Parameter-level mutation.} Many quantum gates require numerical
parameters (\emph{e.g.}, rotational angles in quantum gates $R_x$, $R_y$, and $R_z$). In
parameter-level mutation, \tool replaces such parameters with values sampled
from a predefined set of special constants. These constants include values
historically associated with bugs as well as typical boundary or representative
values, such as $\pm 1$, $0$, and $\pi$. This mutation targets numerical corner
cases while preserving syntactic and semantic validity of the program.

\noindent\textit{Gate-level structural mutation.} Unlike parameter-level mutation, which changes numerical values while preserving circuit topology, gate-level mutation modifies multi-qubit interaction structures. \tool substitutes circuit gates using three classes of rules: (i) entanglement-preserving, (ii) entanglement-expanding, and (iii) entanglement-reducing. These mutations generate structurally different quantum inputs that exercise topology-sensitive logic in compilation, decomposition, simulation, and backend-specific processing. Although such variants may execute already-covered code regions, they can expose failures that are difficult to trigger through parameter mutation alone.

\parh{Test oracle.} \tool utilizes a crash-based oracle to identify runtime anomalies within the executed test cases. To maintain high detection precision, \tool performs exception filtering based on error messages and applies predefined heuristic rules to eliminate common false-positive patterns. The remaining candidate anomalies are then manually inspected to determine if they trigger genuine bugs in the target library. This workflow ensures that \tool identifies impactful bugs while minimizing noise from trivial execution failures.

\section{Experimental Setup}
\subsection{Research Questions}

To enable a comprehensive and rigorous evaluation of the proposed \tool framework, we articulate three key research questions as follows.
\begin{itemize}[leftmargin=*]
    \item[] \textbf{RQ1}: How does \tool perform compared to existing state-of-the-art fuzzing techniques?
    \item[] \textbf{RQ2}: What drives \tool's effectiveness, and how robust is it across different LLM backends and model scales?
    \item[] \textbf{RQ3}:  To what extent can \tool uncover previously unknown bugs in widely used quantum programming libraries?
\end{itemize}

\subsection{Implementation}
We implement a prototype of \tool as a Python-based framework and conduct a multi-faceted evaluation on a high-performance computing platform. All experiments are performed on a Linux server powered by an Intel Xeon Platinum CPU (3.80 GHz, 192 physical cores), 512 GB RAM, and an NVIDIA Tesla H800 GPU, operating under Ubuntu 24.04 LTS. Our evaluation targets three representative quantum computing libraries: Qiskit, PennyLane, and Cirq. The detailed specifications and statistics of these target libraries are summarized in Table~\ref{tab:quantum_libs}. Specifically, the Token column denotes the total number of output tokens generated by Gemini-3~\cite{team2023gemini} during API semantic modeling. 
For seed program generation, we leverage the Ollama platform to deploy the INT8-quantized versions of CodeLlama~\cite{codellama} and Qwen2.5-Coder~\cite{qwen2.5-coder}, balancing inference efficiency with model performance. To maximize hardware throughput and fully utilize the multi-core CPU resources, we implement a multi-processing execution engine that facilitates parallel test execution. Furthermore, to prevent potential resource exhaustion or hangs caused by malformed programs, a strict 20-second execution budget is enforced for each test case.
To evaluate the effectiveness of \tool, we compare it against three state-of-the-art baselines: two quantum-specific fuzzers, MorphQ~\cite{paltenghi2023morphq} and FuzzQ~\cite{shaking2025oopsla}, and one LLM-based fuzzer, Fuzz4All~\cite{xia2024fuzz4all}. A detailed summary of the core hyperparameters, configuration settings, and description of these baselines is provided in SM~C.

\section{Evaluation}
\label{sec:results}

\begin{table}[t]
\centering
\footnotesize
\caption{Overview of mainstream quantum libraries.}
\label{tab:quantum_libs}
\begin{tabular}{lccrrr}
\toprule
\textbf{Library} & \textbf{Version} &  \textbf{GitHub Stars} & \textbf{LoC} & \textbf{API Number} & \textbf{Token} \\
\midrule
Qiskit & 2.3.0 &  7.1K & 45928 & 788 & 0.38M\\
PennyLane & 0.44.0 &  3.1K & 69514 & 1206 & 0.61M\\
Cirq & 1.6.1 &  4.9K & 31774 &  866 & 0.41M\\
\bottomrule
\end{tabular}
\end{table}

\begin{table}[t]
\caption{Comparison with baseline fuzzers on three quantum libraries.}\label{table:test0}
\centering
\small
\setlength{\tabcolsep}{3pt}
\begin{tabular}{@{}c l r r r@{}}
\toprule
\textbf{Target} & \textbf{Fuzzer} & 
\textbf{\# Test Cases}   &
\textbf{Coverage} &
\textbf{\# Unique Crashes} \\
\midrule
\multirow{4}{*}{Qiskit}
 & FuzzQ  & 42985 & 11636(25.34\%) & 5 \\
 & MorphQ   & 25714  & 11396(24.81\%) & 13 \\
 & Fuzz4All & 23478  & 24344(53.00\%) & 339  \\
 & \tool & 35952  & 29078(63.31\%) & 685 \\
\midrule
\multirow{2}{*}{Pennylane}
 & Fuzz4All & 21823  & 31584(45.44\%) & 426 \\
 & \tool & 32068  & 40814(58.71\%) & 1027 \\
\midrule
\multirow{3}{*}{Cirq}
 & FuzzQ  & 27067  & 10027(31.56\%) & 3 \\
 & Fuzz4All & 22547  & 18286(55.35\%) & 391 \\
 & \tool & 34224  & 23446(73.79\%) & 706 \\
\bottomrule
\end{tabular}
\end{table}

\begin{figure}[t]
  \centering
  \includegraphics[width=0.99\columnwidth]{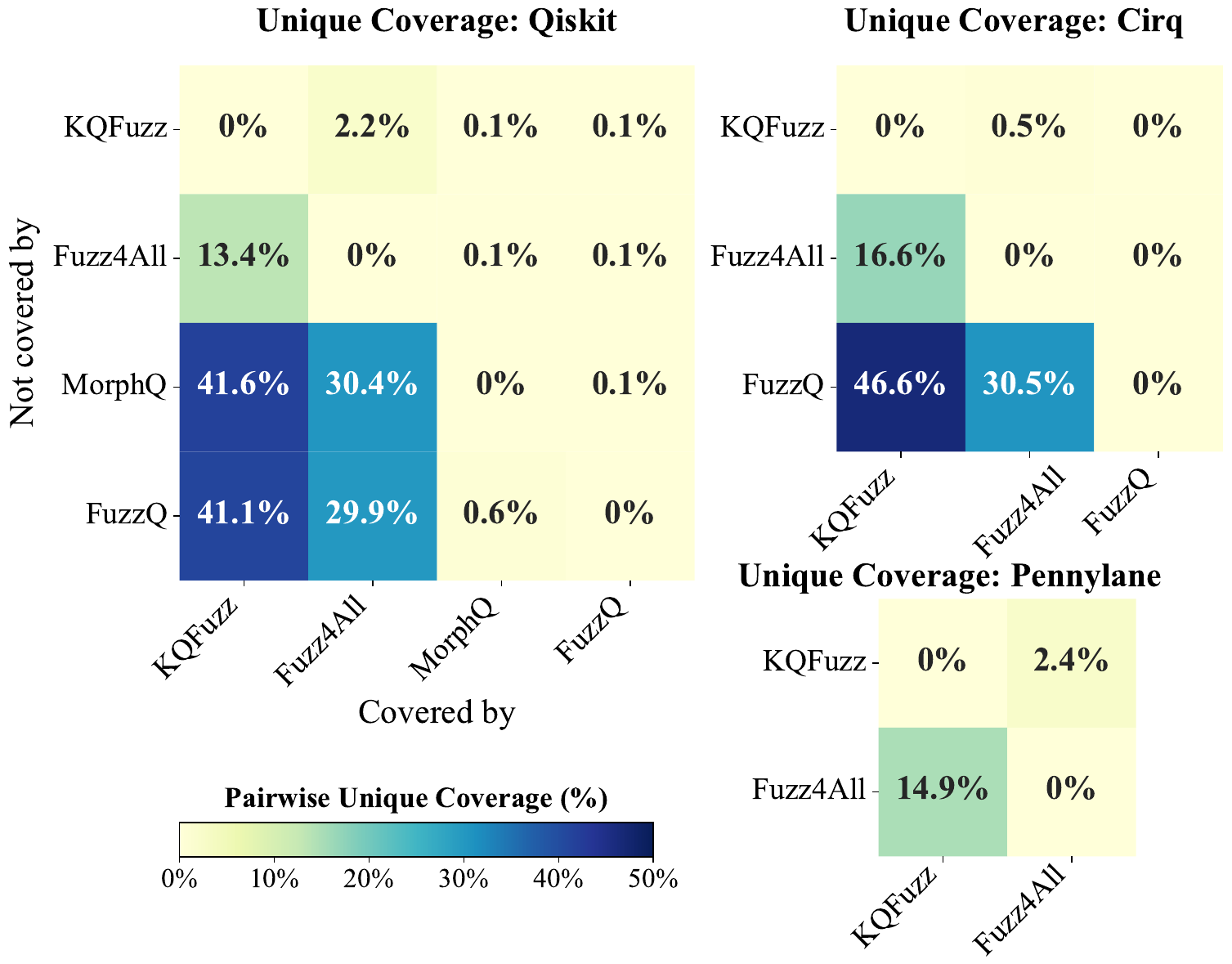}
  \caption{Pairwise unique coverage comparison.}
  \label{fig:heatmap}
\end{figure}

\subsection{Comparison with Other Fuzzers}

To answer RQ1, we compare \tool with each baseline only on targets evaluated by its released implementation: MorphQ on Qiskit, FuzzQ on Qiskit and Cirq, and Fuzz4All on all three libraries. Although the FuzzQ artifact includes a PennyLane adapter, it is presented only as a future-work demonstration rather than an evaluated configuration. We do not port the baselines to additional target libraries, because the resulting performance could be influenced by our implementation choices and might not faithfully represent the original released tools. For every applicable comparison, each fuzzer was executed for a continuous 24-hour period on the same pinned versions (Qiskit 2.3.0, PennyLane 0.44.0, and Cirq 1.6.1), with baseline configurations and hyperparameters set to the defaults recommended in the original papers. Table~\ref{table:test0} summarizes the number
of generated programs, the resulting code coverage, and the number of unique
crashes. Here, we measure Python line coverage using \texttt{coverage.py} while
excluding the libraries' internal test files, and we regard a generated program
as valid only if it executes without runtime exceptions in a properly
configured environment and invokes the target API at least once. To better
compare how much new behavior each fuzzer reaches, Figure~\ref{fig:heatmap}
further visualizes the pairwise unique
coverage relationships between \tool and the baselines. Each cell in the heatmap represents the percentage of the target library's total code lines that are covered by one fuzzer but missed by another.

Experimental results show that \tool consistently achieves the strongest
overall testing effectiveness across all targets. On Qiskit, \tool reaches
63.31\% coverage, outperforming Fuzz4All (53.00\%) and more than doubling the
coverage of MorphQ (24.81\%) and FuzzQ (25.34\%). On PennyLane, \tool
improves coverage from 45.44\% to 58.71\% over Fuzz4All. On Cirq, \tool
achieves 73.79\% coverage, compared with 55.35\% for Fuzz4All and 31.56\% for
FuzzQ. This broader exploration also translates into more bug-finding
opportunities: \tool reports the highest number of unique crashes on every
framework, including 685 on Qiskit, 1027 on PennyLane, and 706 on Cirq.
Moreover, it generates the largest number of programs in all three settings,
indicating that \tool combines higher throughput with stronger exploration. A detailed analysis of the overall time allocation under the default configuration used in RQ1 is provided in SM~\ref{sm:efficiency_overall}.

The pairwise heatmaps in Figure~\ref{fig:heatmap} confirm that \tool's gains represent substantial new behaviors rather than redundant paths. On Qiskit, \tool covers an additional 13.4\% of the total library code that Fuzz4All fails to reach. In contrast, \tool misses only 2.2\% of the library code covered by Fuzz4All. The gap is even more pronounced against specialized quantum fuzzers: MorphQ and FuzzQ fail to cover 41.6\% and 41.1\% of the total library code, respectively, that is otherwise successfully explored by \tool. Similarly, on Cirq, \tool misses only 0.5\% of the code reached by Fuzz4All, while covering an additional 16.6\% of the library. On PennyLane, \tool contributes 14.9\% unique library coverage beyond Fuzz4All, while missing only 2.4\%. These results indicate that \tool largely subsumes the coverage achieved by prior works while significantly expanding the test frontier into previously unreachable regions of the codebases.

\begin{rqbox}
\textbf{Answer to RQ1:} \tool achieves the best results on all three libraries:
63.31\% coverage on Qiskit, 58.71\% on PennyLane, and 73.79\% on Cirq, all above
the strongest baselines, triggering 2.1$\times$ more unique crashes and adding
15.0\% unique coverage on average.
\end{rqbox}

\subsection{Component Contributions and Cross-Model Robustness}
To address RQ2, we first examine how the core components of \tool contribute to its effectiveness and then evaluate whether its advantage over Fuzz4All persists across different LLM backends and model scales.

\parh{Ablation study of core components.}
We form an ablation over Seed Selection and Codebase Knowledge, while the final configuration measures the incremental benefit of adding Mutation to Seed+Repo. Each configuration is evaluated using 6,000 generated test seeds. Table~\ref{tab:ablation_study} reports validity, line coverage, and unique crashes for each configuration.

Codebase Knowledge is the primary contributor to effectiveness. Compared with the base configuration, Repo-only improves validity rate by 8.86--16.70 percentage points (pp), coverage by 6.74--15.52 pp, and unique crashes by 71--262 across the three libraries. For example, on Cirq, it increases coverage from 54.20\% to 69.72\% and validity from 41.37\% to 50.23\%. These results show that library-specific API and constraint information substantially improve the generation of valid and coverage-effective tests.

Seed Selection and Mutation provide complementary benefits. Compared with the base configuration, Seed-only improves validity by 5.58--6.97 pp, while producing smaller coverage and crash gains. Adding Mutation to Seed+Repo improves coverage by only 0.28--0.47 pp, but discovers 57, 111, and 61 additional unique crashes on Qiskit, PennyLane, and Cirq, respectively. Thus, Seed Selection primarily improves seed quality, whereas Mutation mainly expands crash-triggering exploration. Mutation is evaluated on top of Seed+Repo because it is designed to refine selected, codebase-aware seeds rather than operate as an independent generator.

\begin{figure}[t]
  \centering
  \includegraphics[width=0.82\linewidth]{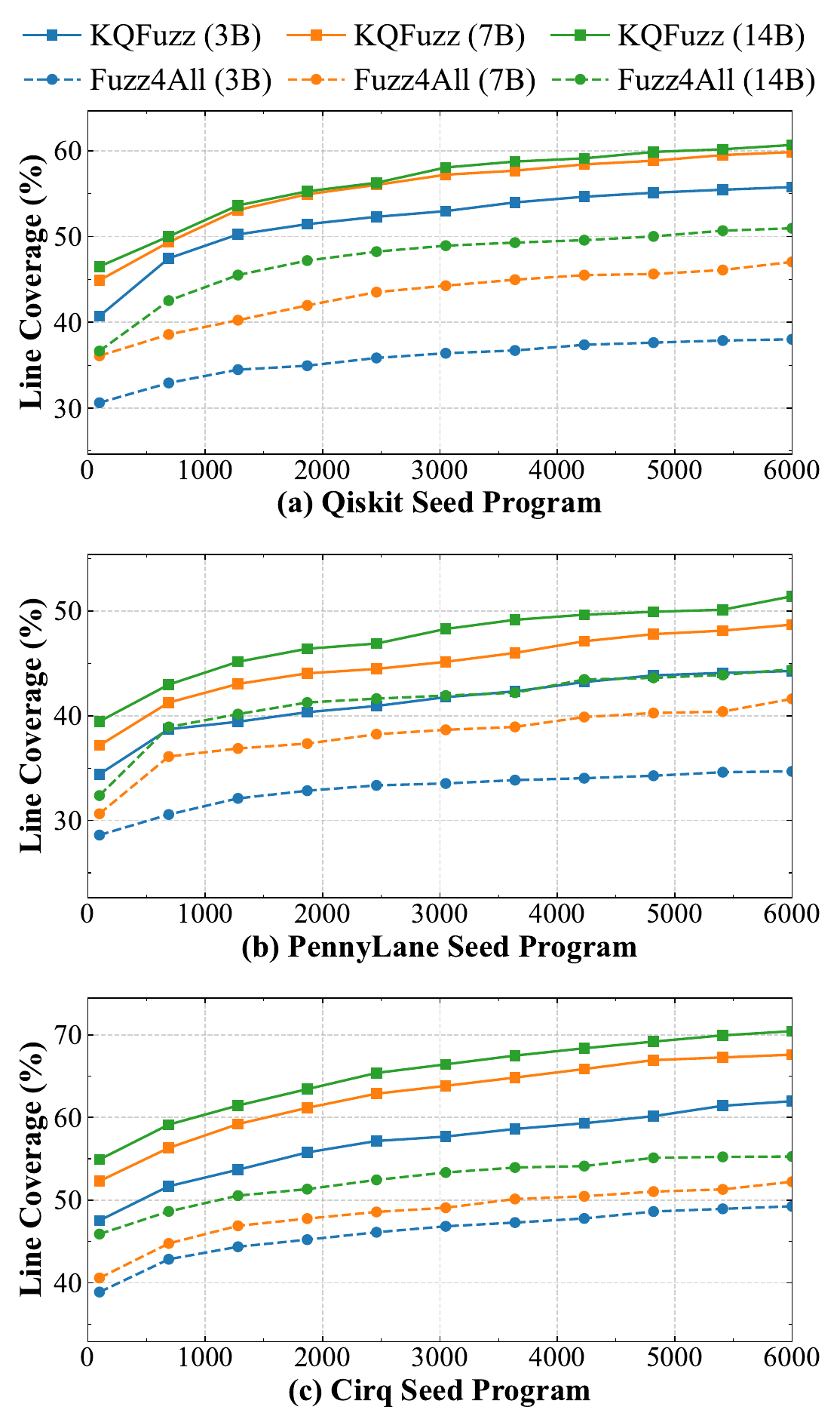}
  \caption{Line coverage comparison.}
  \label{fig:coverage_qwen}
\end{figure}

\begin{table}[t]
\footnotesize
  \centering
  \caption{Ablation results on three quantum libraries.}
  \label{tab:ablation_study}
  \setlength{\tabcolsep}{2pt}
  \begin{tabular}{@{}lcccccc@{}}
    \toprule
    \multirow{2}{*}{\textbf{Target}} & \multicolumn{3}{c}{\textbf{Configuration}} & \multirow{2}{*}{\textbf{\% valid}} & \multirow{2}{*}{\textbf{Coverage}} & \multirow{2}{*}{\textbf{Unique Crashes}} \\
    \cmidrule(lr){2-4}
    & \textbf{Seed} & \textbf{Repo} & \textbf{Mut.} & & & \\
    \midrule
    \multirow{5}{*}{Qiskit} 
    & $\times$     & $\times$     & $\times$     & 33.72\% & 22379(48.73\%) & 155 \\
    & $\checkmark$ & $\times$     & $\times$     & 40.69\% & 23189(50.49\%) & 165(+10) \\
    & $\times$     & $\checkmark$ & $\times$     & 50.42\% & 27945(60.85\%) & 226(+71) \\
    & $\checkmark$ & $\checkmark$ & $\times$     & 53.71\% & 28316(61.65\%) & 250(+95) \\
    & $\checkmark$ & $\checkmark$ & $\checkmark$ & 53.71\% & 28530(62.12\%) & 307(+152) \\
    \midrule
    \multirow{5}{*}{PennyLane}
    & $\times$     & $\times$     & $\times$     & 38.48\% & 33227(47.80\%) & 192 \\
    & $\checkmark$ & $\times$     & $\times$     & 44.06\% & 33632(48.38\%) & 206(+14) \\
    & $\times$     & $\checkmark$ & $\times$     & 49.53\% & 37910(54.54\%) & 454(+262) \\
    & $\checkmark$ & $\checkmark$ & $\times$     & 51.34\% & 38287(54.86\%) & 476(+284) \\
    & $\checkmark$ & $\checkmark$ & $\checkmark$ & 51.34\% & 38330(55.14\%) & 587(+395) \\
    \midrule
    \multirow{5}{*}{Cirq} 
    & $\times$     & $\times$     & $\times$     & 41.37\% & 17222(54.20\%) & 165 \\
    & $\checkmark$ & $\times$     & $\times$     & 48.19\% & 17427(54.85\%) & 179(+14) \\
    & $\times$     & $\checkmark$ & $\times$     & 50.23\% & 22154(69.72\%) & 251(+86) \\
    & $\checkmark$ & $\checkmark$ & $\times$     & 53.08\% & 22540(70.94\%) & 283(+118) \\
    & $\checkmark$ & $\checkmark$ & $\checkmark$ & 53.08\% & 22683(71.39\%) & 344(+179) \\
    \bottomrule
  \end{tabular}
\end{table}

\begin{table}[t]
\footnotesize
\centering
\caption{Comparative analysis of validity and coverage for \tool and Fuzz4All across various LLMs.}
\label{tab:rq2_table}
\begin{tabular}{@{}llcc@{}}
\toprule
\textbf{Model} & \textbf{Method} & \textbf{Val. Rate (\%) $\uparrow$} & \textbf{Total Coverage (Unique)} \\ \midrule
\multirow{2}{*}{Qwen2.5-Coder-3B} & Fuzz4All & 21.04 & 57226 (4162)  \\
 & \tool & \textbf{33.16} (+12.12) & 73079 (20015) \\ \midrule
\multirow{2}{*}{Qwen2.5-Coder-7B} & Fuzz4All & 25.51 & 66866 (6004) \\
 & \tool & \textbf{40.64} (+15.13) & 80203 (19341) \\ \midrule
 \multirow{2}{*}{Qwen2.5-Coder-14B} & Fuzz4All & 28.65 & 71741 (7131) \\
 & \tool & \textbf{52.99} (+24.34) & 82317 (17707) \\ \midrule
\multirow{2}{*}{CodeLlama-7B} & Fuzz4All & 14.42 & 53334 (1968) \\
 & \tool & \textbf{23.90} (+9.48) & 70893 (19527) \\ \midrule
\multirow{2}{*}{CodeLlama-13B} & Fuzz4All & 9.66 & 57300 (3104) \\
 & \tool & \textbf{31.04} (+21.38) & 73364 (19168) \\ \bottomrule
\end{tabular}
\end{table}

\parh{Robustness across LLM backends and scales.}
To further verify the robustness of our framework, we evaluated \tool across two
prominent model families (detailed results for CodeLlama are available in SM~D), and compared
them against the baseline. As illustrated in Figure~\ref{fig:coverage_qwen}, \tool consistently
outperforms the baseline across all tested model scales and library targets. We
observe a striking performance crossover in Qiskit and Cirq, where \tool
utilizing the smallest model scale achieves significantly higher code
coverage than the baseline framework at its largest scale. This massive performance gap indicates that the
effectiveness of \tool is primarily driven by our specialized, library-aware
generation strategy rather than the mere scaling of model parameters.

While both methods tend to benefit from increased model capacity, \tool exhibits higher scaling efficiency. Table~\ref{tab:rq2_table} shows that
\tool maintains a  higher validity rate, with absolute improvements
over the baseline ranging from 9.48 to 24.34 pp. Notably, the advantage of
\tool often becomes more pronounced as the model size increases, with the
validity improvement peaking at 24.34\% on the 14B scale. This suggests that our
framework can better unlock the reasoning potential of larger models. As shown
by the growth curves in Figure~\ref{fig:coverage_qwen}, Fuzz4All coverage is frequently constrained by
a high rate of execution exceptions. In contrast, \tool maintains a higher
validity baseline, which allows larger models to focus more on handling complex
API constraints rather than basic syntax, thereby achieving
higher final coverage. These results demonstrate
that \tool is a robust, model-agnostic framework that can effectively leverage
the latent capabilities of different model families and scales. The corresponding per-sample token consumption and generation time of \tool and Fuzz4All across these model configurations are reported in SM~\ref{sm:efficiency_llm}.

\begin{rqbox}
\textbf{Answer to RQ2:} KQFuzz’s effectiveness is driven by its core components, primarily Codebase Knowledge, which boosts coverage by up to 16.1\% (\emph{e.g.}, on Cirq). Furthermore, KQFuzz exhibits superior scaling efficiency across LLMs, improving validity rates by 9.48\% to 24.34\% and achieving significantly faster coverage convergence.
\end{rqbox}

\begin{table}[t]
\centering
\caption{Bugs found by \tool}
\label{tab:bugs}
\footnotesize
\begin{tabular}{lllc l}
\toprule
\textbf{Target} & \textbf{ID} & \textbf{API Location} & \textbf{Type} & \textbf{Status} \\ 
\midrule
\multirow{6}{*}{Qiskit} 
    & \#15550 & synthesis & BV & Fixed \\
    & \#4159  & synthesis & SV & Confirmed \\
    & \#15657 & circuit   & SD & Fixed \\
    & \#15665 & synthesis & BV & Fixed \\
    & \#15666 & circuit.library & BV & Fixed \\
    & \#15780 & synthesis & BV & Fixed \\ 
\midrule
\multirow{4}{*}{PennyLane} 
    & \#8931  & io.qasm\_interpreter & SD & Fixed \\
    & \#8726  & decomposition & SV & Fixed \\
    & \#9140  & math & SD & Fixed \\
    & \#9141  & transforms & SV & Fixed \\ 
\midrule
\multirow{3}{*}{Cirq} 
    & \#7934  & contrib.paulistring & SD & Fixed \\
    & \#7939  & transformers & SV & Fixed \\
    & \#7941  & neutral\_atoms & SD & Fixed \\ 
\bottomrule
\end{tabular}
\end{table}

\subsection{Bug Detection}
Regarding RQ3, as of the submission date, \tool has identified 13 bugs, with 13 confirmed by developers and 10 already fixed. Table~\ref{tab:bugs} summarizes the identified bugs. We categorize these bugs into three distinct types based on their root causes: (i) \textbf{Boundary violations (BV).} These errors occur when the library fails to handle extreme input dimensions or edge-case configurations. Such bugs typically bypass high-level sanity checks and manifest as low-level system panics, memory exhaustion, or process hangs. (ii) \textbf{State divergence (SD).} This category encompasses defects rooted in flawed internal implementations, organizational structures, or data processing logic. This type of bug usually emerges during multi-stage transformations, where internal metadata fails to maintain synchronized states. (iii) \textbf{Semantic violations (SV).} These defects refer to implementation behaviors that deviate from documented specifications or logical protocols. This includes incorrect gate decomposition logic, inconsistencies between documentation and implementation, or violations of API invariants.

Listing~\ref{code:qiskit_forloop} illustrates a bug case where assigning parameters to a circuit containing \texttt{ForLoopOp} leads to a state divergence, resulting in an inconsistency error. This defect arises from redundant tracking: the loop parameter is
incorrectly registered in the outer circuit's parameter table via both the
control-flow operator slot and the nested circuit body. Such dual-source
registration creates a synchronization gap between the global and local
parameter scopes, causing the assignment mechanism to fail. To understand the
bug, we quote the maintainer's response to our bug report.
\begin{quote}
\textit{Huh, this seems like it should really have been an error on all Qiskit versions, and the error shouldn't get as far as the internal logic error.}
\end{quote}
Their response confirms that this issue has silently existed across all Qiskit
releases and can only be triggered through the specific parameter-reuse pattern
generated by our fuzzer, representing a scenario unlikely to be covered by
existing fuzzers. Additional case studies illustrating other bug categories are provided in SM~D.

\begin{lstlisting}[style=QuantumPyBordered, caption={Nested Parameter Inconsistency in Qiskit}, label={code:qiskit_forloop}]
from qiskit.circuit import QuantumCircuit, Parameter
from qiskit.circuit.controlflow import ForLoopOp

theta = Parameter('theta')
body = QuantumCircuit(1)
body.rx(theta, 0)
for_loop_op = ForLoopOp(range(3), theta, body)
qc = QuantumCircuit(1)
qc.append(for_loop_op, [0])
# This line triggers the error
qc.assign_parameters({theta: 3.14159 / 2}, inplace=True)\end{lstlisting}

\begin{rqbox}
\textbf{Answer to RQ3:} \tool successfully identified 13 bugs across major quantum libraries, with 12 already fixed by developers. By exposing critical boundary, state, and semantic violations that elude conventional testing, our tool demonstrates exceptional capability in ensuring the functional reliability of the evolving quantum libraries.
\end{rqbox}

\section{Threats to Validity}
There are some threats to the validity of our results and the conclusions drawn from them. First, the nondeterministic nature of LLMs could introduce variability in our metrics. We mitigate this threat to validity by running each fuzzer for 24 hours to reduce measurement variability. Second, our crash oracle relies on heuristic exception filtering and deduplication rules to eliminate false positive patterns. This presents a threat because such mechanisms might inadvertently discard genuine bugs that share error signatures with trivial failures, leading to unreported bugs. Finally, our evaluation focuses on three quantum libraries and specific open source models, and performance might differ on other paradigms. We believe our framework can generalize to other platforms that provide sufficient source code transparency for knowledge extraction.

\section{Related Work}
\parh{Quantum software engineering.} The rapid emergence of quantum computing has driven the need for robust software engineering practices across the entire development lifecycle~\cite{zhao2020quantum,ali2022taoyue,murillo2025quantum,oldfield2025faster,wang2024quantum,muqeet2024machine,mendiluze2025quantum,xia2025quantum,ye2025measurement,guo2025m2qcode,jin2025novaq,long2024equivalence,guo2024repairing,zhao2023bugs4q,arcaini2025introduction,muqeet2024mitigating}. Recent advancements have pushed the boundaries of quantum software development, introducing LLM-assisted quantum code synthesis (\emph{e.g.},~\cite{guo2025quanbench}), and specialized compiler optimizations
~\cite{wang2024dac}. Alongside these development tools, quality assurance mechanisms are actively being established to ensure platform reliability. Empirical studies~\cite{paltenghi2022bugs} have characterized unique bug patterns in quantum software, motivating the creation of static analysis tools like Qchecker~\cite{zhao2023qchecker} to detect issues in quantum programs prior to execution. Program-level techniques such as Muskit and QuanFuzz respectively apply syntactically valid mutations and generate test inputs for quantum programs~\cite{muskit2021ase,poster2021icst}. In contrast, QDiff, MorphQ, and FuzzQ target quantum platforms or libraries and use quantum programs as test inputs~\cite{wang2021qdiff,paltenghi2023morphq,shaking2025oopsla}. QEMI adapts EMI to quantum software-stack testing by removing dead code constructed from quantum control-flow patterns and checking the resulting variants for crash and output-distribution inconsistencies~\cite{luo2026qemi}. \tool instead uses source-derived API knowledge to guide program generation toward broad quantum-library API exploration.

\parh{Testing compilers and other developer tools.} 
The critical role of complex software infrastructure has motivated extensive research into automated tool testing. Compiler testing frequently employs randomized code generation and equivalence modulo inputs to uncover optimization bugs. Recently, this scope has expanded to lower-level assemblers via error-driven grammar inference~\cite{kim2024asfuzzer}. Beyond foundational compilers, specialized techniques target diverse input spaces: WebAssembly engines utilize stack-invariant transformations for semantic-preserving mutation~\cite{zhang2025waltzz}, network protocols employ constrained query-response fuzzing to uncover stateful bugs~\cite{zhang2024resolverfuzz}, and database management systems leverage dynamic data-dependency analysis for generic query synthesis~\cite{yang2024buzzbee}. Furthermore, LLMs are revolutionizing the field by acting as universal, language-agnostic fuzzers~\cite{xia2024fuzz4all} and automating the generation of complex API fuzz drivers~\cite{zhang2024effective}.

\section{Conclusion}
This paper presents \tool, a knowledge-guided fuzzing framework for quantum libraries that leverages LLMs to enhance their performance and efficiency. By leveraging a multi-dimensional API knowledge corpus to guide LLM-based seed generation, coupled with fitness-driven seed selection and two-level mutations, \tool successfully overcomes the limitations of insufficient flexibility in traditional circuit-level fuzzers and the low validity rates of generic LLM-based approaches. Experimental evaluations on major quantum libraries, including Qiskit, PennyLane, and Cirq, demonstrate that \tool significantly outperforms state-of-the-art baselines in both code coverage and quantum program validity. Furthermore, the discovery of 13 bugs underscores the framework's practical effectiveness in uncovering critical bugs within the rapidly evolving quantum computing ecosystem.

\section{Data Availability}
The source code and data involved in our study are publicly available in the Zenodo repository~\cite{lk_qfuzz_repo}.

\clearpage
\newpage
\appendix

This supplementary material (\textbf{SM}) provides necessary details omitted from the main text, including quantum computing background (SM~\ref{sm:background}), API association modeling definitions (SM~\ref{sm:api_modeling}), implementation and experimental setup of \tool (SM~\ref{sm:implementation}), additional evaluation results with bug case studies (SM~\ref{sm:results}), and efficiency analysis (SM~\ref{sm:efficiency}).

\section{Background of quantum computing}
\label{sm:background}
This section provides a more comprehensive overview of the fundamental concepts in quantum computing referenced in the main text.

\parh{Qubits and quantum states.}
The fundamental unit of quantum information is the quantum bit, or qubit. Unlike a classical bit, which must reside in a discrete state of either 0 or 1, a qubit can exist in a linear combination, or \emph{superposition}, of these states. Mathematically, a single-qubit state is represented as a vector $|\psi\rangle$ in a two-dimensional complex Hilbert space $\mathcal{H} \cong \mathbb{C}^2$. Using Dirac notation, an arbitrary single-qubit state is expressed as:$$|\psi\rangle = \alpha|0\rangle + \beta|1\rangle,$$where $|0\rangle$ and $|1\rangle$ form the standard computational basis, and $\alpha, \beta \in \mathbb{C}$ are complex probability amplitudes satisfying the normalization condition $|\alpha|^2 + |\beta|^2 = 1$. For an $n$-qubit system, the state space grows exponentially to $2^n$ dimensions, represented by the tensor product of the individual qubit Hilbert spaces ($\mathcal{H}^{\otimes n}$). Within this expanded space, \emph{entanglement} manifests as multi-qubit states that cannot be factored into the tensor product of individual qubit states, representing non-local correlations unique to quantum mechanics.

\parh{Quantum gates.}
The evolution of a closed quantum system is governed by unitary transformations. In the quantum circuit model, these transformations are enacted by quantum gates, which are represented by unitary matrices $U$ satisfying $U^\dagger U = U U^\dagger = I$. Single-qubit gates manipulate the state of individual qubits. Prominent examples include the Pauli matrices ($X, Y, Z$), which induce generalized rotations around the Bloch sphere, and the Hadamard gate ($H$), which creates an equal superposition from a computational basis state:
$$H|0\rangle = \frac{1}{\sqrt{2}}(|0\rangle + |1\rangle).$$
To generate entanglement and enable universal quantum computation, multi-qubit gates are required. A ubiquitous two-qubit entangling operation is the Controlled-NOT (CNOT) gate, which flips the state of a target qubit if and only if the control qubit is in the $|1\rangle$ state. Table~\ref{tab:quantum_gates} summarizes the symbols and mathematical matrix representations of these typical quantum gates. By composing finite sets of single-qubit and entangling two-qubit gates, any arbitrary unitary operation can be approximated to a desired precision.

\begin{table}[htbp]
\centering
\footnotesize
\caption{Summary of typical quantum gates and their matrix representations.}
\label{tab:quantum_gates}
\renewcommand{\arraystretch}{1.2}
\begin{tabular}{@{}lcc@{}} 
\toprule
\textbf{Gate Name} & \textbf{Symbol} & \textbf{Matrix} \\
\midrule
Hadamard & $H$ & $\frac{1}{\sqrt{2}}\begin{pmatrix} 1 & 1 \\ 1 & -1 \end{pmatrix}$ \\
Pauli-X (NOT) & $X$ & $\begin{pmatrix} 0 & 1 \\ 1 & 0 \end{pmatrix}$ \\
Pauli-Y & $Y$ & $\begin{pmatrix} 0 & -i \\ i & 0 \end{pmatrix}$ \\
Pauli-Z & $Z$ & $\begin{pmatrix} 1 & 0 \\ 0 & -1 \end{pmatrix}$ \\
Controlled-NOT & CNOT & $\begin{pmatrix} 1 & 0 & 0 & 0 \\ 0 & 1 & 0 & 0 \\ 0 & 0 & 0 & 1 \\ 0 & 0 & 1 & 0 \end{pmatrix}$ \\
\bottomrule
\end{tabular}
\end{table}

\parh{Measurement.}
To extract classical information from a quantum system, a measurement operation must be applied. According to the postulates of quantum mechanics, measuring a qubit in the computational basis $\{|0\rangle, |1\rangle\}$ forces its superposition state to collapse into one of the definite basis states. This outcome is inherently probabilistic; governed by the Born rule, the probability of observing a specific state is given by the squared magnitude of its corresponding amplitude. For instance, measuring $|\psi\rangle = \alpha|0\rangle + \beta|1\rangle$ yields the outcome $0$ with probability $|\alpha|^2$ and $1$ with probability $|\beta|^2$. Following measurement, the quantum state irreversibly collapses into the observed state, destroying any prior superposition or entanglement.

\parh{Quantum circuits.} A quantum algorithm is practically formulated as a quantum circuit—an ordered, acyclic sequence of operations applied to an initial state, typically initialized to $|0\rangle^{\otimes n}$. Read from left to right, a circuit comprises state preparation, a series of unitary quantum gates to execute the logical computation, and terminal measurements to map the final quantum state into classical registers. Due to the probabilistic nature of quantum measurement, a circuit is typically executed repeatedly to sample the output distribution, thereby allowing the estimation of expectation values that encode the solution to the given computational problem.

\section{Details of API Association Modeling}
\label{sm:api_modeling}
This section elaborates on the mathematical formulations for the API association model referenced in \S~\ref{sec:know_extra} of the main text. To rigorously quantify the semantic and structural relationships between APIs, we formally define the three dimensions used to compute the combined association score.

\parh{Proximity.} We use file- and module-level locality as a proxy for semantic relatedness. Let $S_p(a,b)$ denote the \emph{proximity score} between APIs $a$ and $b$, defined as:
$$S_p(a, b) = \begin{cases} \alpha, & \text{if } a \text{ and } b \text{ are defined in the same file}, \\ \beta, & \text{if } a \text{ and } b \text{ are in the same module but different files}, \\ \lambda, & \text{otherwise.} \end{cases}$$

\parh{Type overlap.} Quantum software typically introduces numerous domain-specific types. APIs that share input or output types are likely to be used together. Let $\text{Typeset}(a)$ denote the set of non-native types appearing in the signature of API $a$. We quantify type-level similarity using the Jaccard index:
$$S_t(a, b) = J(\text{Typeset}(a), \text{Typeset}(b)),$$
where $J(A,B)=|A \cap B| / |A \cup B|$ measures the overlap between two sets.

\parh{Call relationships.} We approximate API relatedness based on call relationships observed in the source code via \emph{reachability} and \emph{commonality}. Reachability is measured using a shortest call distance function $D(a,b)$, where $D(a,b)=1$ indicates a direct call and $D(a,b)=+\infty$ indicates no call path. Commonality captures shared invocation patterns. Let $\text{Caller}(\cdot)$ and $\text{Callee}(\cdot)$ denote the sets of APIs that invoke or are invoked by a given API, respectively. The resulting \emph{call association score} is defined as:
\[
\begin{aligned}
S_c(a,b) = & \frac{1}{2} \left[ J(\text{Caller}(a), \text{Caller}(b)) + J(\text{Callee}(a), \text{Callee}(b)) \right] \\
& + \frac{1}{2D(a,b)}.
\end{aligned}
\]

\section{Implementation details of \tool}
\label{sm:implementation}
This section elaborates on the implementation details and the experimental setup of \tool omitted in \S~\ref{sec:approach}. Table~\ref{tab:hyperparams} provides a comprehensive summary of the core hyperparameters and configuration settings used throughout the evaluation of \tool. Subsequently, we describe the exact prompt structures designed for \tool and define the primary metrics utilized to evaluate performance. 

\begin{table}[h]
\centering
\caption{Summary of Core Hyperparameters and Configuration Settings.}
\label{tab:hyperparams}
\footnotesize
\begin{tabular}{llcc}
\toprule
\textbf{Section} & \textbf{Parameter} & \textbf{Symbol} & \textbf{Value} \\ \midrule
\multirow{5}{*}{\S~\ref{sec:know_extra}} & Intra-file Score & $\alpha$ & 1.0 \\
 & Intra-module Score & $\beta$ & 0.6 \\
 & Default Proximity Score & $\lambda$ & 0 \\ 
 & Evolution Sensitivity & $\gamma$ & 4 \\
 & Evolution Range & $\delta$ & 0.3 \\ \midrule
\multirow{3}{*}{\S~\ref{sec:know_extra}} & Locality Weight & $w_1$ & 2 \\
 & Type Overlap Weight & $w_2$ & 1 \\
 & Call Relationship Weight & $w_3$ & 2 \\ \midrule
\S~\ref{sec:seed_selection} & Entanglement Cap & $\tau$ & 6 \\ 
& API Diversity Weight & $\eta$ & 3 \\ \midrule
\multirow{2}{*}{\S~\ref{sec:seed_selection}} & Selection Top-$\kappa$ & $\kappa$ & 10 \\
 & Programs per Iteration & - & 30 \\ \bottomrule
\end{tabular}
\end{table}

\begin{figure}[t]
  \centering
  \includegraphics[width=0.9\columnwidth]{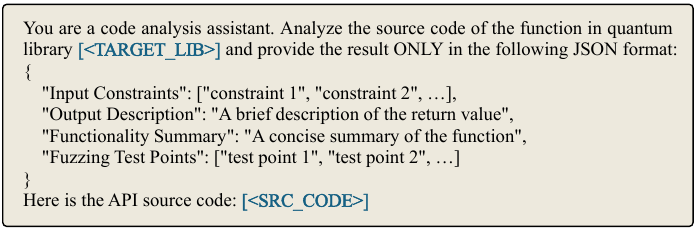}
  \caption{Prompt structure used by \tool for API implementation semantic modeling.}
  \label{fig:prompt_1}
\end{figure}

\begin{figure}[t]
  \centering
  \includegraphics[width=0.94\columnwidth]{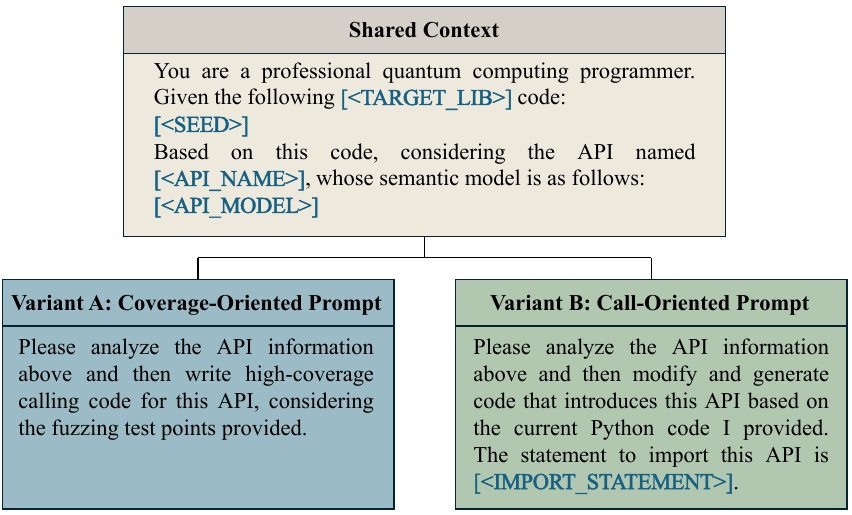}
  \caption{A structured view of the prompt template used for LLM-guided seed program generation. The template consists of a shared context block (seed program and target API information) and two parallel, specialized variant blocks: coverage-oriented prompt and call-oriented prompt.}
  \label{fig:prompt}
\end{figure}

\subsection{Prompt}

In this section, we detail the specific prompt structures utilized by \tool, corresponding to the two main phases of our framework: API semantic extraction and seed program generation.

\parh{API implementation semantic modeling prompt.}
As illustrated in Figure~\ref{fig:prompt_1}, to isolate the exact logic of the target library, the prompt injects the library name (\texttt{<TARGET\_LIB>}) and the relevant, filtered source code (\texttt{<SRC\_CODE>}). To ensure the output is easily parsed by the automated testing pipeline, we constrain the LLM to return a strictly formatted JSON object. This JSON explicitly constructs the semantic model by extracting:
\begin{itemize}
    \item \emph{Input Constraints:} rules, types, and boundary conditions for the API's arguments;
    \item \emph{Output Description:} expected return values and behaviors;
    \item \emph{Functionality Summary:} a concise overview of the API's purpose; and
    \item \emph{Fuzzing Test Points:} specific edge cases, parameter combinations, and structural targets to guide the fuzzing phase.
\end{itemize}

\parh{LLM-guided seed generation prompt.}
Figure~\ref{fig:prompt} displays the prompt template used by the \emph{fuzzing LLM}. To maintain contextual awareness and ensure the generation of valid seed programs, the prompt is divided into a \textbf{Shared Context} block and two task-specific variants:
\begin{itemize}
    \item \emph{Shared Context:} this foundational block establishes the persona of a professional quantum computing programmer. It grounds the generation process by providing the execution skeleton via the seed program (\texttt{<SEED>}), alongside the target library (\texttt{<TARGET\_LIB>}), the target API name (\texttt{<API\_NAME>}), and its comprehensive semantic model \\(\texttt{<API\_MODEL>}) extracted during the modeling phase;
    \item \emph{Variant A (Coverage-Oriented Prompt):} activated when the target API ($i_t$) is already present in the seed program. It directs the model to analyze the semantic profile and generate high-coverage calling code, specifically instructing the LLM to leverage the provided fuzzing test points to explore deeper, more diverse execution paths; and
    \item \emph{Variant B (Call-Oriented Prompt):} activated when $i_t$ is absent from the seed program. It instructs the model to modify the existing Python seed to seamlessly introduce a new invocation of the target API. To guarantee syntactic correctness and a fully runnable output, this variant explicitly injects the necessary import statement (\texttt{<IMPORT\_STATEMENT>}).
\end{itemize}

\subsection{Metric}

We adopt three primary metrics in software testing to analyze the collected experimental results. For clarity, their key concepts and functionalities are introduced below.

\parh{Code coverage.} Code coverage has been widely adopted in software testing and quantum library testing. We measure Python line coverage using the \texttt{coverage.py} tool while excluding the library’s internal test files to ensure that the metrics accurately reflect the exploration of core functional logic. Furthermore, we evaluate \textbf{unique code coverage}, defined as the specific code segments triggered exclusively by a particular configuration. This metric allows us to quantify the distinct exploratory effectiveness of each setting and its unique contribution to the overall code space exploration.

\parh{Validity.} A generated case is defined as valid if it executes without runtime exceptions in a properly configured environment and invokes the target API at least once. We perform deduplication to count only unique instances, which further provides metrics such as the number of valid programs and validity rate. Note that because the validity of mutated samples inherently depends on the specific mutation strategy applied, mutants are excluded from the statistics for this metric.

\parh{Bug detection.} Following prior work on quantum library fuzzing, we report the number of unique detected bugs.

\subsection{Baseline}
To evaluate the effectiveness of \tool, we compared it against three state-of-the-art baselines, including two quantum-specific fuzzers and one LLM-based universal fuzzer:

\begin{itemize}
    \item \textbf{MorphQ}~\cite{paltenghi2023morphq} is a metamorphic testing framework specifically developed for the Qiskit platform, which designs quantum-specific metamorphic relations and a dedicated program generator to expose semantic inconsistencies.
    \item \textbf{FuzzQ}~\cite{shaking2025oopsla} encodes QASM semantics in Alloy to generate structurally constrained circuits and employs invariant checking, statistical tests, and cross-simulator unitary consistency as differential oracles.
    \item \textbf{Fuzz4All}~\cite{xia2024fuzz4all} is a general-purpose fuzzer that proposes an LLM-based auto-prompting mechanism combined with an iterative fuzzing loop to automatically synthesize diverse and semantically meaningful inputs across programming languages, achieving high coverage and broad applicability beyond quantum systems.
\end{itemize}

\section{Experimental results}
\label{sm:results}

This section provides supplementary experimental data and qualitative examples that complement the primary evaluation presented in \S~\ref{sec:results} of the main text. Specifically, \S~\ref{sm:coverage} presents additional code coverage results utilizing the CodeLlama model family to reinforce our primary coverage findings. Furthermore, as referenced in the main text, \S~\ref{sm:case_study} details additional case studies illustrating other bug categories discovered by \tool.

\subsection{Coverage Results}
\label{sm:coverage}
\begin{figure}[t]
  \centering
  \includegraphics[width=0.8\linewidth]{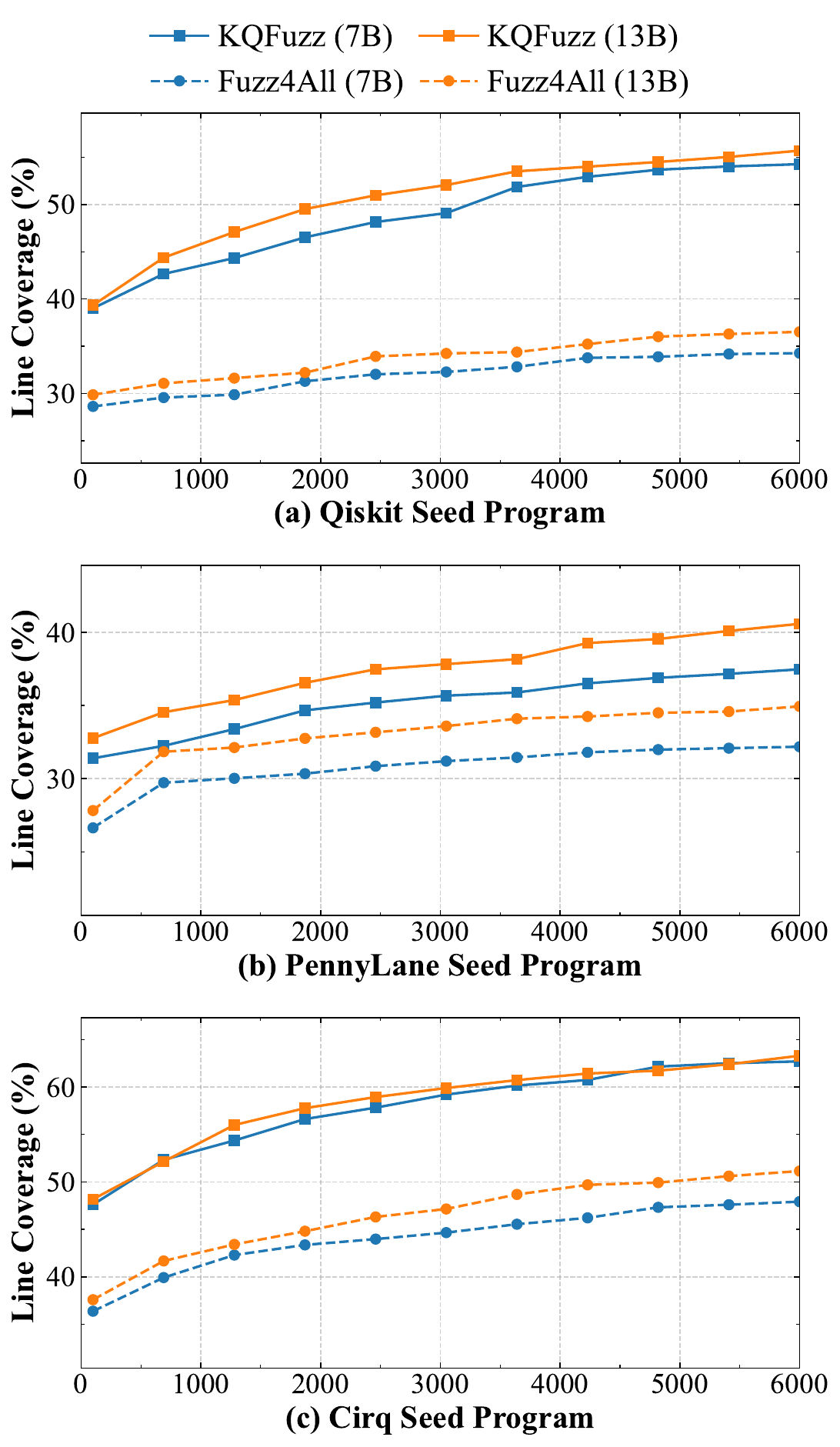}
  \caption{Line coverage comparison using CodeLlama.}
  \label{fig:coverage_codellama}
\end{figure}

As illustrated in Figure~\ref{fig:coverage_codellama}, the evaluation utilizing the CodeLlama model family (7B and 13B) corroborates the findings presented in the main text. \tool consistently outperforms the Fuzz4All baseline across all three quantum libraries. Most notably, \tool operating at the smaller 7B scale consistently achieves significantly higher line coverage than the baseline framework utilizing the larger 13B model throughout the entire generation process. These supplementary results further validate that the superiority of \tool is fundamentally driven by its meticulously designed, library-aware generation strategy rather than the sheer scaling of model parameters.

\subsection{Case Study}
\label{sm:case_study}
In the following, we showcase several representative bugs found by \tool to illustrate their impact and root causes.

Listing~\ref{code:qiskit_boundary} shows a boundary violation where synthesizing a 1-qubit Clifford operator triggers a Rust-level panic. Although the LNN (Linear Nearest Neighbor) synthesis should support any valid circuit dimension, the underlying logic failed to handle the $n=1$ edge cases, assuming multi-qubit connectivity. It is confirmed as an unhandled boundary in the synthesis engine and has since been patched to support minimal qubit configurations.

\begin{lstlisting}[style=QuantumPyBordered, caption={Single-qubit Clifford Synthesis Panic in Qiskit}, label={code:qiskit_boundary}]
from qiskit.circuit import QuantumCircuit
from qiskit.quantum_info import Clifford
from qiskit.synthesis.clifford import synth_clifford_depth_lnn

qc = QuantumCircuit(1)
qc.h(0)
clifford_op = Clifford(qc)

# This call triggers the Rust panic
synthesized_circuit = synth_clifford_depth_lnn(clifford_op)\end{lstlisting}

Listing~\ref{code:cirq_hash} illustrates a state divergence where checking gate compatibility triggers a \texttt{TypeError}. The callee function fails when processing a \texttt{UniformSuperpositionGate} because the gate class is unhashable. This bug stems from a flaw in the relevant \texttt{\_\_contains\_\_} implementation, which incorrectly assumes that all gate-derived objects are hashable.

\begin{lstlisting}[style=QuantumPyBordered, caption={Unhashable Gate Membership in Cirq}, label={code:cirq_hash}]
import cirq
from cirq.neutral_atoms import is_native_neutral_atom_gate
from cirq.ops import UniformSuperpositionGate

gate = UniformSuperpositionGate(m_value=3, num_qubits=2)

# This call triggers the TypeError
print(is_native_neutral_atom_gate(gate))\end{lstlisting}

Listing~\ref{code:pennylane_sign_expand} illustrates a semantic violation where executing a QNode with the \texttt{sign\_expand} transform results in an error. This defect stems from a packaging oversight where a critical metadata file (\texttt{sign\_expand\_data.json}) was omitted from the distribution configuration. This missing resource prevents the transformation API from loading its required logic, resulting in a runtime failure that violates the library's functional integrity and documented behavior.
\begin{lstlisting}[style=QuantumPyBordered, caption={Missing Resource Dependency in PennyLane}, label={code:pennylane_sign_expand}]
import pennylane as qml
dev = qml.device("default.qubit", wires=2)
obs = qml.Hamiltonian([2.0, -1.5], [qml.PauliZ(0), qml.PauliX(1)])

@qml.transforms.sign_expand
@qml.qnode(dev)
def circuit():
    qml.RX(0.5, wires=0)
    qml.RY(0.5, wires=1)
    qml.CNOT(wires=[0, 1])
    return qml.expval(obs)
# A FileNotFoundError error occurs during circuit execution.
result = circuit()\end{lstlisting}

\section{Efficiency Analysis}
\label{sm:efficiency}

This section analyzes the computational efficiency of \tool. First, we examine the one-time modeling overhead and the execution-time distribution of the complete pipeline under the default configuration used in RQ1. Second, we report the per-sample token consumption and generation time across the LLM backends evaluated in RQ2.

\subsection{Overall Computational Efficiency}
\label{sm:efficiency_overall}

\parh{Cost of the modeling LLM.}
The modeling LLM is invoked exclusively during the one-time API corpus
construction phase, in which the source code of each API is processed once to
construct its semantic model. The resulting corpus is reused throughout all
subsequent fuzzing iterations without any further invocation of the modeling
LLM.
As reported in Table~\ref{tab:quantum_libs} of the main text, this phase
consumes 0.38M, 0.61M, and 0.41M tokens for Qiskit, PennyLane, and Cirq,
respectively. Corpus construction takes only 2.3, 3.3, and 1.6 minutes for the
three libraries, corresponding to 0.15\%, 0.23\%, and 0.11\% of the fixed
24-hour budget. Therefore, the one-time preprocessing overhead is negligible
in all three cases.

\parh{Time breakdown across the pipeline.}
To examine where computational time is spent, we divide the complete \tool
pipeline into four stages: Preparation, Seed Generation, Mutation, and
Execution. Preparation corresponds to the one-time API corpus construction.
Seed Generation invokes the fuzzing LLM to produce seed programs. Mutation
performs source-level transformations on existing programs, whereas Execution
runs the generated programs against the target library.

\begin{table}[t]
\centering
\caption{Execution-Time Distribution.}
\label{tab:pipeline_time}
\footnotesize
\setlength{\tabcolsep}{3pt}
\begin{tabular}{lrrrr}
\toprule
\textbf{Target} &
\textbf{Preparation} &
\textbf{Seed Gen.} &
\textbf{Mutation} &
\textbf{Execution} \\
\midrule
Qiskit    & 0.15\% & 57.50\% & 0.13\% & 42.22\% \\
PennyLane & 0.23\% & 50.83\% & 0.15\% & 48.79\% \\
Cirq      & 0.11\% & 55.97\% & 0.14\% & 43.78\% \\
\bottomrule
\end{tabular}
\end{table}

As shown in Table~\ref{tab:pipeline_time}, the computational cost is almost
entirely concentrated in Seed Generation and Execution, which together account
for more than 99\% of the total execution time across all three libraries. In
contrast, Preparation and Mutation each consume less than 0.23\%. Preparation
is negligible because corpus construction is performed only once and reused
throughout all subsequent iterations, while Mutation is implemented as a
lightweight source-level transformation without LLM invocation. This design
allows \tool to efficiently diversify LLM-generated seed programs under the
fixed experimental budget.

\subsection{Per-Sample Generation Cost}
\label{sm:efficiency_llm}

We further analyze the cost of individual LLM generation calls across the
model families and scales evaluated in RQ2. For both \tool and Fuzz4All, we
report the average input tokens, output tokens, and generation time per sample.
Each result is averaged over Qiskit, PennyLane, and Cirq.

\begin{table}[t]
\centering
\caption{Average Per-Sample Token Consumption and Generation Time.}
\label{tab:per_sample_efficiency}
\footnotesize
\setlength{\tabcolsep}{3pt}
\begin{tabular}{llrrr}
\toprule
\textbf{Model} &
\textbf{Method} &
\textbf{Input Tokens} &
\textbf{Output Tokens} &
\textbf{Time (s)} \\
\midrule
\multirow{2}{*}{Qwen-3B}
 & Fuzz4All & 332.15 & 201.43 & 0.92 \\
 & \tool    & 367.10 & 300.90 & 1.32 \\
\midrule
\multirow{2}{*}{Qwen-7B}
 & Fuzz4All & 365.40 & 245.56 & 1.44 \\
 & \tool    & 329.38 & 257.33 & 1.52 \\
\midrule
\multirow{2}{*}{Qwen-14B}
 & Fuzz4All & 323.67 & 193.33 & 1.89 \\
 & \tool    & 385.60 & 344.00 & 3.35 \\
\midrule
\multirow{2}{*}{CodeLlama-7B}
 & Fuzz4All & 518.21 & 318.73 & 1.48 \\
 & \tool    & 479.17 & 431.50 & 2.07 \\
\midrule
\multirow{2}{*}{CodeLlama-13B}
 & Fuzz4All & 526.95 & 319.67 & 2.47 \\
 & \tool    & 521.84 & 479.50 & 3.85 \\
\bottomrule
\end{tabular}
\end{table}

As shown in Table~\ref{tab:per_sample_efficiency}, both methods incur similar input token consumption, while the clear and consistent difference lies in output tokens, where KQFuzz consumes more than Fuzz4All on every model. This is a direct consequence of KQFuzz's generation paradigm: rather than generating short and standalone programs, KQFuzz iteratively extends existing seed programs by introducing new and semantically related API calls, resulting in longer and structurally richer programs. This design is precisely what enables KQFuzz to achieve significantly higher coverage and validity, and the additional output tokens represent meaningful semantic content rather than redundancy.


\begin{thebibliography}{68}


\ifx \showCODEN    \undefined \def \showCODEN     #1{\unskip}     \fi
\ifx \showDOI      \undefined \def \showDOI       #1{#1}\fi
\ifx \showISBNx    \undefined \def \showISBNx     #1{\unskip}     \fi
\ifx \showISBNxiii \undefined \def \showISBNxiii  #1{\unskip}     \fi
\ifx \showISSN     \undefined \def \showISSN      #1{\unskip}     \fi
\ifx \showLCCN     \undefined \def \showLCCN      #1{\unskip}     \fi
\ifx \shownote     \undefined \def \shownote      #1{#1}          \fi
\ifx \showarticletitle \undefined \def \showarticletitle #1{#1}   \fi
\ifx \showURL      \undefined \def \showURL       {\relax}        \fi
\providecommand\bibfield[2]{#2}
\providecommand\bibinfo[2]{#2}
\providecommand\natexlab[1]{#1}
\providecommand\showeprint[2][]{arXiv:#2}

\bibitem[Abbas et~al\mbox{.}(2024)]%
        {abbas2024nrp}
\bibfield{author}{\bibinfo{person}{Amira Abbas}, \bibinfo{person}{Andris
  Ambainis}, \bibinfo{person}{Brandon Augustino}, \bibinfo{person}{Andreas
  B{\"a}rtschi}, \bibinfo{person}{Harry Buhrman}, \bibinfo{person}{Carleton
  Coffrin}, \bibinfo{person}{Giorgio Cortiana}, \bibinfo{person}{Vedran
  Dunjko}, \bibinfo{person}{Daniel~J Egger}, \bibinfo{person}{Bruce~G
  Elmegreen}, {et~al\mbox{.}}} \bibinfo{year}{2024}\natexlab{}.
\newblock \showarticletitle{Challenges and opportunities in quantum
  optimization}.
\newblock \bibinfo{journal}{\emph{Nature Reviews Physics}} \bibinfo{volume}{6},
  \bibinfo{number}{12} (\bibinfo{year}{2024}), \bibinfo{pages}{718--735}.
\newblock


\bibitem[Ali et~al\mbox{.}(2022)]%
        {ali2022taoyue}
\bibfield{author}{\bibinfo{person}{Shaukat Ali}, \bibinfo{person}{Tao Yue},
  {and} \bibinfo{person}{Rui Abreu}.} \bibinfo{year}{2022}\natexlab{}.
\newblock \showarticletitle{When software engineering meets quantum computing}.
\newblock \bibinfo{journal}{\emph{Commun. ACM}} \bibinfo{volume}{65},
  \bibinfo{number}{4} (\bibinfo{year}{2022}), \bibinfo{pages}{84--88}.
\newblock


\bibitem[{Anonymous Authors}(2026)]%
        {lk_qfuzz_repo}
\bibfield{author}{\bibinfo{person}{{Anonymous Authors}}.}
  \bibinfo{year}{2026}\natexlab{}.
\newblock \bibinfo{title}{{Replication package for anonymous submission}}.
\newblock
\newblock
\urldef\tempurl%
\url{https://doi.org/10.5281/zenodo.19230459}
\showDOI{\tempurl}


\bibitem[Arcaini et~al\mbox{.}(2025)]%
        {arcaini2025introduction}
\bibfield{author}{\bibinfo{person}{Paolo Arcaini}, \bibinfo{person}{Andriy
  Miranskyy}, {and} \bibinfo{person}{Hausi M{\"u}ller}.}
  \bibinfo{year}{2025}\natexlab{}.
\newblock \bibinfo{title}{Introduction to the Special Section on software
  engineering for hybrid quantum computing systems}.
\newblock , \bibinfo{numpages}{112362}~pages.
\newblock


\bibitem[Assolini et~al\mbox{.}(2024)]%
        {assolini2024sas}
\bibfield{author}{\bibinfo{person}{Nicola Assolini},
  \bibinfo{person}{Alessandra Di~Pierro}, {and} \bibinfo{person}{Isabella
  Mastroeni}.} \bibinfo{year}{2024}\natexlab{}.
\newblock \showarticletitle{Static analysis of quantum programs}. In
  \bibinfo{booktitle}{\emph{International Static Analysis Symposium}}.
  Springer, \bibinfo{pages}{1--25}.
\newblock


\bibitem[Bergholm et~al\mbox{.}(2018)]%
        {pennylane}
\bibfield{author}{\bibinfo{person}{Ville Bergholm}, \bibinfo{person}{Josh
  Izaac}, \bibinfo{person}{Maria Schuld}, \bibinfo{person}{Christian Gogolin},
  \bibinfo{person}{Shahnawaz Ahmed}, \bibinfo{person}{Vishnu Ajith},
  \bibinfo{person}{M~Sohaib Alam}, \bibinfo{person}{Guillermo Alonso-Linaje},
  \bibinfo{person}{Bharath AkashNarayanan}, \bibinfo{person}{Ali Asadi},
  {et~al\mbox{.}}} \bibinfo{year}{2018}\natexlab{}.
\newblock \showarticletitle{Pennylane: Automatic differentiation of hybrid
  quantum-classical computations}.
\newblock \bibinfo{journal}{\emph{arXiv preprint arXiv:1811.04968}}
  (\bibinfo{year}{2018}).
\newblock


\bibitem[B{\"o}hme et~al\mbox{.}(2020)]%
        {bohme2020boosting}
\bibfield{author}{\bibinfo{person}{Marcel B{\"o}hme},
  \bibinfo{person}{Valentin~JM Man{\`e}s}, {and} \bibinfo{person}{Sang~Kil
  Cha}.} \bibinfo{year}{2020}\natexlab{}.
\newblock \showarticletitle{Boosting fuzzer efficiency: An information
  theoretic perspective}. In \bibinfo{booktitle}{\emph{Proceedings of the 28th
  ACM Joint Meeting on European Software Engineering Conference and Symposium
  on the Foundations of Software Engineering}}. \bibinfo{pages}{678--689}.
\newblock


\bibitem[B{\"o}hme et~al\mbox{.}(2017)]%
        {bohme2017directed}
\bibfield{author}{\bibinfo{person}{Marcel B{\"o}hme},
  \bibinfo{person}{Van-Thuan Pham}, \bibinfo{person}{Manh-Dung Nguyen}, {and}
  \bibinfo{person}{Abhik Roychoudhury}.} \bibinfo{year}{2017}\natexlab{}.
\newblock \showarticletitle{Directed greybox fuzzing}. In
  \bibinfo{booktitle}{\emph{Proceedings of the 2017 ACM SIGSAC conference on
  computer and communications security}}. \bibinfo{pages}{2329--2344}.
\newblock


\bibitem[Cerezo et~al\mbox{.}(2021)]%
        {cerezo2021nrp}
\bibfield{author}{\bibinfo{person}{Marco Cerezo}, \bibinfo{person}{Andrew
  Arrasmith}, \bibinfo{person}{Ryan Babbush}, \bibinfo{person}{Simon~C
  Benjamin}, \bibinfo{person}{Suguru Endo}, \bibinfo{person}{Keisuke Fujii},
  \bibinfo{person}{Jarrod~R McClean}, \bibinfo{person}{Kosuke Mitarai},
  \bibinfo{person}{Xiao Yuan}, \bibinfo{person}{Lukasz Cincio},
  {et~al\mbox{.}}} \bibinfo{year}{2021}\natexlab{}.
\newblock \showarticletitle{Variational quantum algorithms}.
\newblock \bibinfo{journal}{\emph{Nature Reviews Physics}} \bibinfo{volume}{3},
  \bibinfo{number}{9} (\bibinfo{year}{2021}), \bibinfo{pages}{625--644}.
\newblock


\bibitem[{Cirq Developers}(2025)]%
        {cirq_developers_2025_10.5281/zenodo.4062499}
\bibfield{author}{\bibinfo{person}{{Cirq Developers}}.}
  \bibinfo{year}{2025}\natexlab{}.
\newblock \bibinfo{booktitle}{\emph{Cirq}}.
\newblock
\urldef\tempurl%
\url{https://doi.org/10.5281/zenodo.4062499}
\showDOI{\tempurl}


\bibitem[Deng et~al\mbox{.}(2023)]%
        {deng2023issta}
\bibfield{author}{\bibinfo{person}{Yinlin Deng},
  \bibinfo{person}{Chunqiu~Steven Xia}, \bibinfo{person}{Haoran Peng},
  \bibinfo{person}{Chenyuan Yang}, {and} \bibinfo{person}{Lingming Zhang}.}
  \bibinfo{year}{2023}\natexlab{}.
\newblock \showarticletitle{Large language models are zero-shot fuzzers:
  Fuzzing deep-learning libraries via large language models}. In
  \bibinfo{booktitle}{\emph{Proceedings of the 32nd ACM SIGSOFT international
  symposium on software testing and analysis}}. \bibinfo{pages}{423--435}.
\newblock


\bibitem[Deng et~al\mbox{.}(2024)]%
        {deng2024icse}
\bibfield{author}{\bibinfo{person}{Yinlin Deng},
  \bibinfo{person}{Chunqiu~Steven Xia}, \bibinfo{person}{Chenyuan Yang},
  \bibinfo{person}{Shizhuo~Dylan Zhang}, \bibinfo{person}{Shujing Yang}, {and}
  \bibinfo{person}{Lingming Zhang}.} \bibinfo{year}{2024}\natexlab{}.
\newblock \showarticletitle{Large language models are edge-case generators:
  Crafting unusual programs for fuzzing deep learning libraries}. In
  \bibinfo{booktitle}{\emph{Proceedings of the 46th IEEE/ACM international
  conference on software engineering}}. \bibinfo{pages}{1--13}.
\newblock


\bibitem[Dong et~al\mbox{.}(2025)]%
        {dong2025tdsc}
\bibfield{author}{\bibinfo{person}{Ruiqi Dong}, \bibinfo{person}{Fanke Tong},
  \bibinfo{person}{He Huang}, \bibinfo{person}{Xiaogang Zhu},
  \bibinfo{person}{Xi Xiao}, \bibinfo{person}{Shaohua Wang},
  \bibinfo{person}{Sheng Wen}, {and} \bibinfo{person}{Yang Xiang}.}
  \bibinfo{year}{2025}\natexlab{}.
\newblock \showarticletitle{One Mutation Fits All: Exploring Universal Library
  Fuzzing based on Exogenous Mutation}.
\newblock \bibinfo{journal}{\emph{IEEE Transactions on Dependable and Secure
  Computing}} (\bibinfo{year}{2025}).
\newblock


\bibitem[Fang and Ying(2024)]%
        {fang2024pldi}
\bibfield{author}{\bibinfo{person}{Wang Fang} {and} \bibinfo{person}{Mingsheng
  Ying}.} \bibinfo{year}{2024}\natexlab{}.
\newblock \showarticletitle{Symbolic execution for quantum error correction
  programs}.
\newblock \bibinfo{journal}{\emph{Proceedings of the ACM on Programming
  Languages}} \bibinfo{volume}{8}, \bibinfo{number}{PLDI}
  (\bibinfo{year}{2024}), \bibinfo{pages}{1040--1065}.
\newblock


\bibitem[Fioraldi et~al\mbox{.}(2020)]%
        {fioraldi2020afl}
\bibfield{author}{\bibinfo{person}{Andrea Fioraldi}, \bibinfo{person}{Dominik
  Maier}, \bibinfo{person}{Heiko Ei{\ss}feldt}, {and} \bibinfo{person}{Marc
  Heuse}.} \bibinfo{year}{2020}\natexlab{}.
\newblock \showarticletitle{$\{$AFL++$\}$: Combining incremental steps of
  fuzzing research}. In \bibinfo{booktitle}{\emph{14th USENIX workshop on
  offensive technologies (WOOT 20)}}.
\newblock


\bibitem[Fowler et~al\mbox{.}(2012)]%
        {fowler2012surface}
\bibfield{author}{\bibinfo{person}{Austin~G Fowler}, \bibinfo{person}{Matteo
  Mariantoni}, \bibinfo{person}{John~M Martinis}, {and}
  \bibinfo{person}{Andrew~N Cleland}.} \bibinfo{year}{2012}\natexlab{}.
\newblock \showarticletitle{Surface codes: Towards practical large-scale
  quantum computation}.
\newblock \bibinfo{journal}{\emph{Physical Review A—Atomic, Molecular, and
  Optical Physics}} \bibinfo{volume}{86}, \bibinfo{number}{3}
  (\bibinfo{year}{2012}), \bibinfo{pages}{032324}.
\newblock


\bibitem[Gill et~al\mbox{.}(2025)]%
        {gill2025quantumc}
\bibfield{author}{\bibinfo{person}{Sukhpal~Singh Gill}, \bibinfo{person}{Oktay
  Cetinkaya}, \bibinfo{person}{Stefano Marrone}, \bibinfo{person}{Daniel
  Claudino}, \bibinfo{person}{David Haunschild}, \bibinfo{person}{Leon
  Schlote}, \bibinfo{person}{Huaming Wu}, \bibinfo{person}{Carlo Ottaviani},
  \bibinfo{person}{Xiaoyuan Liu}, \bibinfo{person}{Sree~Pragna Machupalli},
  {et~al\mbox{.}}} \bibinfo{year}{2025}\natexlab{}.
\newblock \showarticletitle{Quantum computing: Vision and challenges}.
\newblock In \bibinfo{booktitle}{\emph{Quantum computing}}.
  \bibinfo{publisher}{Elsevier}, \bibinfo{pages}{19--42}.
\newblock


\bibitem[Gill et~al\mbox{.}(2022)]%
        {quantum}
\bibfield{author}{\bibinfo{person}{Sukhpal~Singh Gill}, \bibinfo{person}{Adarsh
  Kumar}, \bibinfo{person}{Harvinder Singh}, \bibinfo{person}{Manmeet Singh},
  \bibinfo{person}{Kamalpreet Kaur}, \bibinfo{person}{Muhammad Usman}, {and}
  \bibinfo{person}{Rajkumar Buyya}.} \bibinfo{year}{2022}\natexlab{}.
\newblock \showarticletitle{Quantum computing: A taxonomy, systematic review
  and future directions}.
\newblock \bibinfo{journal}{\emph{Software: Practice and Experience}}
  \bibinfo{volume}{52}, \bibinfo{number}{1} (\bibinfo{year}{2022}),
  \bibinfo{pages}{66--114}.
\newblock


\bibitem[Guo et~al\mbox{.}(2025a)]%
        {guo2025m2qcode}
\bibfield{author}{\bibinfo{person}{Xiaoyu Guo}, \bibinfo{person}{Shinobu
  Saito}, {and} \bibinfo{person}{Jianjun Zhao}.}
  \bibinfo{year}{2025}\natexlab{a}.
\newblock \showarticletitle{M2QCode: A Model-Driven Framework for Generating
  Multi-Platform Quantum Programs}.
\newblock \bibinfo{journal}{\emph{arXiv preprint arXiv:2510.17110}}
  (\bibinfo{year}{2025}).
\newblock


\bibitem[Guo et~al\mbox{.}(2025b)]%
        {guo2025quanbench}
\bibfield{author}{\bibinfo{person}{Xiaoyu Guo}, \bibinfo{person}{Minggu Wang},
  {and} \bibinfo{person}{Jianjun Zhao}.} \bibinfo{year}{2025}\natexlab{b}.
\newblock \showarticletitle{QuanBench: Benchmarking Quantum Code Generation
  with Large Language Models}.
\newblock \bibinfo{journal}{\emph{arXiv preprint arXiv:2510.16779}}
  (\bibinfo{year}{2025}).
\newblock


\bibitem[Guo et~al\mbox{.}(2024)]%
        {guo2024repairing}
\bibfield{author}{\bibinfo{person}{Xiaoyu Guo}, \bibinfo{person}{Jianjun Zhao},
  {and} \bibinfo{person}{Pengzhan Zhao}.} \bibinfo{year}{2024}\natexlab{}.
\newblock \showarticletitle{On repairing quantum programs using ChatGPT}. In
  \bibinfo{booktitle}{\emph{Proceedings of the 5th ACM/IEEE International
  Workshop on Quantum Software Engineering}}. \bibinfo{pages}{9--16}.
\newblock


\bibitem[Hu et~al\mbox{.}(2024)]%
        {hu2024issta}
\bibfield{author}{\bibinfo{person}{Tianmin Hu}, \bibinfo{person}{Guixin Ye},
  \bibinfo{person}{Zhanyong Tang}, \bibinfo{person}{Shin~Hwei Tan},
  \bibinfo{person}{Huanting Wang}, \bibinfo{person}{Meng Li}, {and}
  \bibinfo{person}{Zheng Wang}.} \bibinfo{year}{2024}\natexlab{}.
\newblock \showarticletitle{Upbeat: Test input checks of q\# quantum
  libraries}. In \bibinfo{booktitle}{\emph{Proceedings of the 33rd ACM SIGSOFT
  International Symposium on Software Testing and Analysis}}.
  \bibinfo{pages}{186--198}.
\newblock


\bibitem[Hui et~al\mbox{.}(2024)]%
        {qwen2.5-coder}
\bibfield{author}{\bibinfo{person}{Binyuan Hui}, \bibinfo{person}{Jian Yang},
  \bibinfo{person}{Zeyu Cui}, \bibinfo{person}{Jiaxi Yang},
  \bibinfo{person}{Dayiheng Liu}, \bibinfo{person}{Lei Zhang},
  \bibinfo{person}{Tianyu Liu}, \bibinfo{person}{Jiajun Zhang},
  \bibinfo{person}{Bowen Yu}, \bibinfo{person}{Keming Lu}, {et~al\mbox{.}}}
  \bibinfo{year}{2024}\natexlab{}.
\newblock \showarticletitle{Qwen2. 5-coder technical report}.
\newblock \bibinfo{journal}{\emph{arXiv preprint arXiv:2409.12186}}
  (\bibinfo{year}{2024}).
\newblock


\bibitem[Javadi-Abhari et~al\mbox{.}(2024)]%
        {qiskit}
\bibfield{author}{\bibinfo{person}{Ali Javadi-Abhari}, \bibinfo{person}{Matthew
  Treinish}, \bibinfo{person}{Kevin Krsulich}, \bibinfo{person}{Christopher~J
  Wood}, \bibinfo{person}{Jake Lishman}, \bibinfo{person}{Julien Gacon},
  \bibinfo{person}{Simon Martiel}, \bibinfo{person}{Paul~D Nation},
  \bibinfo{person}{Lev~S Bishop}, \bibinfo{person}{Andrew~W Cross},
  {et~al\mbox{.}}} \bibinfo{year}{2024}\natexlab{}.
\newblock \showarticletitle{Quantum computing with Qiskit}.
\newblock \bibinfo{journal}{\emph{arXiv preprint arXiv:2405.08810}}
  (\bibinfo{year}{2024}).
\newblock


\bibitem[Jiang et~al\mbox{.}(2024)]%
        {jiang2024fuzzing}
\bibfield{author}{\bibinfo{person}{Yu Jiang}, \bibinfo{person}{Jie Liang},
  \bibinfo{person}{Fuchen Ma}, \bibinfo{person}{Yuanliang Chen},
  \bibinfo{person}{Chijin Zhou}, \bibinfo{person}{Yuheng Shen},
  \bibinfo{person}{Zhiyong Wu}, \bibinfo{person}{Jingzhou Fu},
  \bibinfo{person}{Mingzhe Wang}, \bibinfo{person}{Shanshan Li},
  {et~al\mbox{.}}} \bibinfo{year}{2024}\natexlab{}.
\newblock \showarticletitle{When fuzzing meets llms: Challenges and
  opportunities}. In \bibinfo{booktitle}{\emph{Companion Proceedings of the
  32nd ACM International Conference on the Foundations of Software
  Engineering}}. \bibinfo{pages}{492--496}.
\newblock


\bibitem[Jin et~al\mbox{.}(2025)]%
        {jin2025novaq}
\bibfield{author}{\bibinfo{person}{Tiancheng Jin}, \bibinfo{person}{Shangzhou
  Xia}, {and} \bibinfo{person}{Jianjun Zhao}.} \bibinfo{year}{2025}\natexlab{}.
\newblock \showarticletitle{NovaQ: Improving Quantum Program Testing through
  Diversity-Guided Test Case Generation}.
\newblock \bibinfo{journal}{\emph{arXiv preprint arXiv:2509.04763}}
  (\bibinfo{year}{2025}).
\newblock


\bibitem[Kim et~al\mbox{.}(2024)]%
        {kim2024asfuzzer}
\bibfield{author}{\bibinfo{person}{Hyungseok Kim}, \bibinfo{person}{Soomin
  Kim}, \bibinfo{person}{Jungwoo Lee}, {and} \bibinfo{person}{Sang~Kil Cha}.}
  \bibinfo{year}{2024}\natexlab{}.
\newblock \showarticletitle{AsFuzzer: Differential testing of assemblers with
  error-driven grammar inference}. In \bibinfo{booktitle}{\emph{Proceedings of
  the 33rd ACM SIGSOFT International Symposium on Software Testing and
  Analysis}}. \bibinfo{pages}{1099--1111}.
\newblock


\bibitem[Klimis et~al\mbox{.}(2025)]%
        {shaking2025oopsla}
\bibfield{author}{\bibinfo{person}{Vasileios Klimis}, \bibinfo{person}{Avner
  Bensoussan}, \bibinfo{person}{Elena Chachkarova}, \bibinfo{person}{Karine
  Even-Mendoza}, \bibinfo{person}{Sophie Fortz}, {and} \bibinfo{person}{Connor
  Lenihan}.} \bibinfo{year}{2025}\natexlab{}.
\newblock \showarticletitle{Shaking Up Quantum Simulators with Fuzzing and
  Rigour}.
\newblock \bibinfo{journal}{\emph{Proceedings of the ACM on Programming
  Languages}} \bibinfo{volume}{9}, \bibinfo{number}{OOPSLA2}
  (\bibinfo{year}{2025}), \bibinfo{pages}{1400--1428}.
\newblock


\bibitem[Lewis et~al\mbox{.}(2020)]%
        {lewis2020retrieval}
\bibfield{author}{\bibinfo{person}{Patrick Lewis}, \bibinfo{person}{Ethan
  Perez}, \bibinfo{person}{Aleksandra Piktus}, \bibinfo{person}{Fabio Petroni},
  \bibinfo{person}{Vladimir Karpukhin}, \bibinfo{person}{Naman Goyal},
  \bibinfo{person}{Heinrich K{\"u}ttler}, \bibinfo{person}{Mike Lewis},
  \bibinfo{person}{Wen-tau Yih}, \bibinfo{person}{Tim Rockt{\"a}schel},
  {et~al\mbox{.}}} \bibinfo{year}{2020}\natexlab{}.
\newblock \showarticletitle{Retrieval-augmented generation for
  knowledge-intensive nlp tasks}.
\newblock \bibinfo{journal}{\emph{Advances in neural information processing
  systems}}  \bibinfo{volume}{33} (\bibinfo{year}{2020}),
  \bibinfo{pages}{9459--9474}.
\newblock


\bibitem[Li et~al\mbox{.}(2026)]%
        {li2026methodological}
\bibfield{author}{\bibinfo{person}{Yuechen Li}, \bibinfo{person}{Minqi Shao},
  \bibinfo{person}{Jianjun Zhao}, {and} \bibinfo{person}{Qichen Wang}.}
  \bibinfo{year}{2026}\natexlab{}.
\newblock \showarticletitle{A Methodological Analysis of Empirical Studies in
  Quantum Software Testing}.
\newblock \bibinfo{journal}{\emph{arXiv preprint arXiv:2601.08367}}
  (\bibinfo{year}{2026}).
\newblock


\bibitem[Lin et~al\mbox{.}(2025)]%
        {lin2025ase}
\bibfield{author}{\bibinfo{person}{Xingshuang Lin}, \bibinfo{person}{Qinge
  Xie}, \bibinfo{person}{Binbin Zhao}, \bibinfo{person}{Yuan Tian},
  \bibinfo{person}{Saman Zonouz}, \bibinfo{person}{Na Ruan},
  \bibinfo{person}{Jiliang Li}, \bibinfo{person}{Raheem Beyah}, {and}
  \bibinfo{person}{Shouling Ji}.} \bibinfo{year}{2025}\natexlab{}.
\newblock \showarticletitle{PROMFUZZ: Leveraging LLM-Driven and Bug-Oriented
  Composite Analysis for Detecting Functional Bugs in Smart Contracts}.
\newblock \bibinfo{journal}{\emph{arXiv preprint arXiv:2503.23718}}
  (\bibinfo{year}{2025}).
\newblock


\bibitem[Long and Zhao(2024a)]%
        {long2024equivalence}
\bibfield{author}{\bibinfo{person}{Peixun Long} {and} \bibinfo{person}{Jianjun
  Zhao}.} \bibinfo{year}{2024}\natexlab{a}.
\newblock \showarticletitle{Equivalence, identity, and unitarity checking in
  black-box testing of quantum programs}.
\newblock \bibinfo{journal}{\emph{Journal of Systems and Software}}
  \bibinfo{volume}{211} (\bibinfo{year}{2024}), \bibinfo{pages}{112000}.
\newblock


\bibitem[Long and Zhao(2024b)]%
        {long2024tosem}
\bibfield{author}{\bibinfo{person}{Peixun Long} {and} \bibinfo{person}{Jianjun
  Zhao}.} \bibinfo{year}{2024}\natexlab{b}.
\newblock \showarticletitle{Testing multi-subroutine quantum programs: From
  unit testing to integration testing}.
\newblock \bibinfo{journal}{\emph{ACM Transactions on Software Engineering and
  Methodology}} \bibinfo{volume}{33}, \bibinfo{number}{6}
  (\bibinfo{year}{2024}), \bibinfo{pages}{1--61}.
\newblock


\bibitem[Luo et~al\mbox{.}(2026)]%
        {luo2026qemi}
\bibfield{author}{\bibinfo{person}{Junjie Luo}, \bibinfo{person}{Shangzhou
  Xia}, \bibinfo{person}{Fuyuan Zhang}, {and} \bibinfo{person}{Jianjun Zhao}.}
  \bibinfo{year}{2026}\natexlab{}.
\newblock \showarticletitle{QEMI: A Quantum Software Stacks Testing Framework
  via Equivalence Modulo Inputs}. In \bibinfo{booktitle}{\emph{International
  Conference on Fundamental Approaches to Software Engineering}}. Springer,
  \bibinfo{pages}{149--169}.
\newblock


\bibitem[Man{\`e}s et~al\mbox{.}(2019)]%
        {manes2019tse}
\bibfield{author}{\bibinfo{person}{Valentin~JM Man{\`e}s},
  \bibinfo{person}{HyungSeok Han}, \bibinfo{person}{Choongwoo Han},
  \bibinfo{person}{Sang~Kil Cha}, \bibinfo{person}{Manuel Egele},
  \bibinfo{person}{Edward~J Schwartz}, {and} \bibinfo{person}{Maverick Woo}.}
  \bibinfo{year}{2019}\natexlab{}.
\newblock \showarticletitle{The art, science, and engineering of fuzzing: A
  survey}.
\newblock \bibinfo{journal}{\emph{IEEE Transactions on Software Engineering}}
  \bibinfo{volume}{47}, \bibinfo{number}{11} (\bibinfo{year}{2019}),
  \bibinfo{pages}{2312--2331}.
\newblock


\bibitem[Mendiluze et~al\mbox{.}(2021)]%
        {muskit2021ase}
\bibfield{author}{\bibinfo{person}{E{\~n}aut Mendiluze},
  \bibinfo{person}{Shaukat Ali}, \bibinfo{person}{Paolo Arcaini}, {and}
  \bibinfo{person}{Tao Yue}.} \bibinfo{year}{2021}\natexlab{}.
\newblock \showarticletitle{Muskit: A mutation analysis tool for quantum
  software testing}. In \bibinfo{booktitle}{\emph{2021 36th IEEE/ACM
  International Conference on Automated Software Engineering (ASE)}}. IEEE,
  \bibinfo{pages}{1266--1270}.
\newblock


\bibitem[Mendiluze~Usandizaga et~al\mbox{.}(2025)]%
        {mendiluze2025quantum}
\bibfield{author}{\bibinfo{person}{E{\~n}aut Mendiluze~Usandizaga},
  \bibinfo{person}{Shaukat Ali}, \bibinfo{person}{Tao Yue}, {and}
  \bibinfo{person}{Paolo Arcaini}.} \bibinfo{year}{2025}\natexlab{}.
\newblock \showarticletitle{Quantum circuit mutants: Empirical analysis and
  recommendations}.
\newblock \bibinfo{journal}{\emph{Empirical Software Engineering}}
  \bibinfo{volume}{30}, \bibinfo{number}{4} (\bibinfo{year}{2025}),
  \bibinfo{pages}{100}.
\newblock


\bibitem[Muqeet et~al\mbox{.}(2024a)]%
        {muqeet2024machine}
\bibfield{author}{\bibinfo{person}{Asmar Muqeet}, \bibinfo{person}{Shaukat
  Ali}, \bibinfo{person}{Tao Yue}, {and} \bibinfo{person}{Paolo Arcaini}.}
  \bibinfo{year}{2024}\natexlab{a}.
\newblock \showarticletitle{A machine learning-based error mitigation approach
  for reliable software development on IBM’s quantum computers}. In
  \bibinfo{booktitle}{\emph{Companion Proceedings of the 32nd ACM International
  Conference on the Foundations of Software Engineering}}.
  \bibinfo{pages}{80--91}.
\newblock


\bibitem[Muqeet et~al\mbox{.}(2024b)]%
        {muqeet2024mitigating}
\bibfield{author}{\bibinfo{person}{Asmar Muqeet}, \bibinfo{person}{Tao Yue},
  \bibinfo{person}{Shaukat Ali}, {and} \bibinfo{person}{Paolo Arcaini}.}
  \bibinfo{year}{2024}\natexlab{b}.
\newblock \showarticletitle{Mitigating noise in quantum software testing using
  machine learning}.
\newblock \bibinfo{journal}{\emph{IEEE Transactions on Software Engineering}}
  \bibinfo{volume}{50}, \bibinfo{number}{11} (\bibinfo{year}{2024}),
  \bibinfo{pages}{2947--2961}.
\newblock


\bibitem[Murillo et~al\mbox{.}(2025)]%
        {murillo2025quantum}
\bibfield{author}{\bibinfo{person}{Juan~Manuel Murillo}, \bibinfo{person}{Jose
  Garcia-Alonso}, \bibinfo{person}{Enrique Moguel}, \bibinfo{person}{Johanna
  Barzen}, \bibinfo{person}{Frank Leymann}, \bibinfo{person}{Shaukat Ali},
  \bibinfo{person}{Tao Yue}, \bibinfo{person}{Paolo Arcaini},
  \bibinfo{person}{Ricardo P{\'e}rez-Castillo}, \bibinfo{person}{Ignacio
  Garc{\'\i}a-Rodr{\'\i}guez~de Guzm{\'a}n}, {et~al\mbox{.}}}
  \bibinfo{year}{2025}\natexlab{}.
\newblock \showarticletitle{Quantum software engineering: Roadmap and
  challenges ahead}.
\newblock \bibinfo{journal}{\emph{ACM Transactions on Software Engineering and
  Methodology}} \bibinfo{volume}{34}, \bibinfo{number}{5}
  (\bibinfo{year}{2025}), \bibinfo{pages}{1--48}.
\newblock


\bibitem[Oldfield et~al\mbox{.}(2025)]%
        {oldfield2025faster}
\bibfield{author}{\bibinfo{person}{Noah~H Oldfield}, \bibinfo{person}{Christoph
  Laaber}, \bibinfo{person}{Tao Yue}, {and} \bibinfo{person}{Shaukat Ali}.}
  \bibinfo{year}{2025}\natexlab{}.
\newblock \showarticletitle{Faster and better quantum software testing through
  specification reduction and projective measurements}.
\newblock \bibinfo{journal}{\emph{ACM Transactions on Software Engineering and
  Methodology}} \bibinfo{volume}{34}, \bibinfo{number}{7}
  (\bibinfo{year}{2025}), \bibinfo{pages}{1--39}.
\newblock


\bibitem[Paltenghi and Pradel(2022)]%
        {paltenghi2022bugs}
\bibfield{author}{\bibinfo{person}{Matteo Paltenghi} {and}
  \bibinfo{person}{Michael Pradel}.} \bibinfo{year}{2022}\natexlab{}.
\newblock \showarticletitle{Bugs in quantum computing platforms: an empirical
  study}.
\newblock \bibinfo{journal}{\emph{Proceedings of the ACM on Programming
  Languages}} \bibinfo{volume}{6}, \bibinfo{number}{OOPSLA1}
  (\bibinfo{year}{2022}), \bibinfo{pages}{1--27}.
\newblock


\bibitem[Paltenghi and Pradel(2023)]%
        {paltenghi2023morphq}
\bibfield{author}{\bibinfo{person}{Matteo Paltenghi} {and}
  \bibinfo{person}{Michael Pradel}.} \bibinfo{year}{2023}\natexlab{}.
\newblock \showarticletitle{MorphQ: Metamorphic testing of the Qiskit quantum
  computing platform}. In \bibinfo{booktitle}{\emph{2023 IEEE/ACM 45th
  International Conference on Software Engineering (ICSE)}}. IEEE,
  \bibinfo{pages}{2413--2424}.
\newblock


\bibitem[Roziere et~al\mbox{.}(2023)]%
        {codellama}
\bibfield{author}{\bibinfo{person}{Baptiste Roziere}, \bibinfo{person}{Jonas
  Gehring}, \bibinfo{person}{Fabian Gloeckle}, \bibinfo{person}{Sten Sootla},
  \bibinfo{person}{Itai Gat}, \bibinfo{person}{Xiaoqing~Ellen Tan},
  \bibinfo{person}{Yossi Adi}, \bibinfo{person}{Jingyu Liu},
  \bibinfo{person}{Romain Sauvestre}, \bibinfo{person}{Tal Remez},
  {et~al\mbox{.}}} \bibinfo{year}{2023}\natexlab{}.
\newblock \showarticletitle{Code llama: Open foundation models for code}.
\newblock \bibinfo{journal}{\emph{arXiv preprint arXiv:2308.12950}}
  (\bibinfo{year}{2023}).
\newblock


\bibitem[Shafiuzzaman et~al\mbox{.}(2024)]%
        {staticanalysis}
\bibfield{author}{\bibinfo{person}{Md Shafiuzzaman}, \bibinfo{person}{Achintya
  Desai}, \bibinfo{person}{Laboni Sarker}, {and} \bibinfo{person}{Tevfik
  Bultan}.} \bibinfo{year}{2024}\natexlab{}.
\newblock \showarticletitle{STASE: Static analysis guided symbolic execution
  for UEFI vulnerability signature generation}. In
  \bibinfo{booktitle}{\emph{Proceedings of the 39th IEEE/ACM International
  Conference on Automated Software Engineering}}. \bibinfo{pages}{1783--1794}.
\newblock


\bibitem[She et~al\mbox{.}(2024)]%
        {she2024ccs}
\bibfield{author}{\bibinfo{person}{Dongdong She}, \bibinfo{person}{Adam
  Storek}, \bibinfo{person}{Yuchong Xie}, \bibinfo{person}{Seoyoung Kweon},
  \bibinfo{person}{Prashast Srivastava}, {and} \bibinfo{person}{Suman Jana}.}
  \bibinfo{year}{2024}\natexlab{}.
\newblock \showarticletitle{Fox: Coverage-guided fuzzing as online stochastic
  control}. In \bibinfo{booktitle}{\emph{Proceedings of the 2024 on ACM SIGSAC
  Conference on Computer and Communications Security}}.
  \bibinfo{pages}{765--779}.
\newblock


\bibitem[Stephens et~al\mbox{.}(2016)]%
        {stephens2016driller}
\bibfield{author}{\bibinfo{person}{Nick Stephens}, \bibinfo{person}{John
  Grosen}, \bibinfo{person}{Christopher Salls}, \bibinfo{person}{Andrew
  Dutcher}, \bibinfo{person}{Ruoyu Wang}, \bibinfo{person}{Jacopo Corbetta},
  \bibinfo{person}{Yan Shoshitaishvili}, \bibinfo{person}{Christopher Kruegel},
  {and} \bibinfo{person}{Giovanni Vigna}.} \bibinfo{year}{2016}\natexlab{}.
\newblock \showarticletitle{Driller: Augmenting fuzzing through selective
  symbolic execution.}. In \bibinfo{booktitle}{\emph{NDSS}},
  Vol.~\bibinfo{volume}{16}. \bibinfo{pages}{1--16}.
\newblock


\bibitem[Team et~al\mbox{.}(2023)]%
        {team2023gemini}
\bibfield{author}{\bibinfo{person}{Gemini Team}, \bibinfo{person}{Rohan Anil},
  \bibinfo{person}{Sebastian Borgeaud}, \bibinfo{person}{Jean-Baptiste
  Alayrac}, \bibinfo{person}{Jiahui Yu}, \bibinfo{person}{Radu Soricut},
  \bibinfo{person}{Johan Schalkwyk}, \bibinfo{person}{Andrew~M Dai},
  \bibinfo{person}{Anja Hauth}, \bibinfo{person}{Katie Millican},
  {et~al\mbox{.}}} \bibinfo{year}{2023}\natexlab{}.
\newblock \showarticletitle{Gemini: a family of highly capable multimodal
  models}.
\newblock \bibinfo{journal}{\emph{arXiv preprint arXiv:2312.11805}}
  (\bibinfo{year}{2023}).
\newblock


\bibitem[Upadhyay et~al\mbox{.}(2026)]%
        {upadhyay2026understandingbugsquantumsimulators}
\bibfield{author}{\bibinfo{person}{Krishna Upadhyay}, \bibinfo{person}{Moshood
  Fakorede}, {and} \bibinfo{person}{Umar Farooq}.}
  \bibinfo{year}{2026}\natexlab{}.
\newblock \bibinfo{title}{Understanding Bugs in Quantum Simulators: An
  Empirical Study}.
\newblock
\newblock
\showeprint[arxiv]{2603.22789}~[quant-ph]
\urldef\tempurl%
\url{https://arxiv.org/abs/2603.22789}
\showURL{%
\tempurl}


\bibitem[Wang et~al\mbox{.}(2024b)]%
        {wang2024dac}
\bibfield{author}{\bibinfo{person}{Hanrui Wang}, \bibinfo{person}{Daniel~Bochen
  Tan}, \bibinfo{person}{Pengyu Liu}, \bibinfo{person}{Yilian Liu},
  \bibinfo{person}{Jiaqi Gu}, \bibinfo{person}{Jason Cong}, {and}
  \bibinfo{person}{Song Han}.} \bibinfo{year}{2024}\natexlab{b}.
\newblock \showarticletitle{Q-pilot: Field programmable qubit array compilation
  with flying ancillas}. In \bibinfo{booktitle}{\emph{Proceedings of the 61st
  ACM/IEEE Design Automation Conference}}. \bibinfo{pages}{1--6}.
\newblock


\bibitem[Wang et~al\mbox{.}(2021a)]%
        {poster2021icst}
\bibfield{author}{\bibinfo{person}{Jiyuan Wang}, \bibinfo{person}{Fucheng Ma},
  {and} \bibinfo{person}{Yu Jiang}.} \bibinfo{year}{2021}\natexlab{a}.
\newblock \showarticletitle{Poster: Fuzz testing of quantum program}. In
  \bibinfo{booktitle}{\emph{2021 14th IEEE Conference on Software Testing,
  Verification and Validation (ICST)}}. IEEE, \bibinfo{pages}{466--469}.
\newblock


\bibitem[Wang et~al\mbox{.}(2021b)]%
        {wang2021qdiff}
\bibfield{author}{\bibinfo{person}{Jiyuan Wang}, \bibinfo{person}{Qian Zhang},
  \bibinfo{person}{Guoqing~Harry Xu}, {and} \bibinfo{person}{Miryung Kim}.}
  \bibinfo{year}{2021}\natexlab{b}.
\newblock \showarticletitle{QDiff: Differential testing of quantum software
  stacks}. In \bibinfo{booktitle}{\emph{2021 36th IEEE/ACM international
  conference on automated software engineering (ASE)}}. IEEE,
  \bibinfo{pages}{692--704}.
\newblock


\bibitem[Wang et~al\mbox{.}(2024a)]%
        {wang2024quantum}
\bibfield{author}{\bibinfo{person}{Xinyi Wang}, \bibinfo{person}{Shaukat Ali},
  \bibinfo{person}{Tao Yue}, {and} \bibinfo{person}{Paolo Arcaini}.}
  \bibinfo{year}{2024}\natexlab{a}.
\newblock \showarticletitle{Quantum approximate optimization algorithm for test
  case optimization}.
\newblock \bibinfo{journal}{\emph{IEEE Transactions on Software Engineering}}
  \bibinfo{volume}{50}, \bibinfo{number}{12} (\bibinfo{year}{2024}),
  \bibinfo{pages}{3249--3264}.
\newblock


\bibitem[Wei et~al\mbox{.}(2022)]%
        {cot}
\bibfield{author}{\bibinfo{person}{Jason Wei}, \bibinfo{person}{Xuezhi Wang},
  \bibinfo{person}{Dale Schuurmans}, \bibinfo{person}{Maarten Bosma},
  \bibinfo{person}{Fei Xia}, \bibinfo{person}{Ed Chi}, \bibinfo{person}{Quoc~V
  Le}, \bibinfo{person}{Denny Zhou}, {et~al\mbox{.}}}
  \bibinfo{year}{2022}\natexlab{}.
\newblock \showarticletitle{Chain-of-thought prompting elicits reasoning in
  large language models}.
\newblock \bibinfo{journal}{\emph{Advances in neural information processing
  systems}}  \bibinfo{volume}{35} (\bibinfo{year}{2022}),
  \bibinfo{pages}{24824--24837}.
\newblock


\bibitem[Wu et~al\mbox{.}(2022)]%
        {wu2022one}
\bibfield{author}{\bibinfo{person}{Mingyuan Wu}, \bibinfo{person}{Ling Jiang},
  \bibinfo{person}{Jiahong Xiang}, \bibinfo{person}{Yanwei Huang},
  \bibinfo{person}{Heming Cui}, \bibinfo{person}{Lingming Zhang}, {and}
  \bibinfo{person}{Yuqun Zhang}.} \bibinfo{year}{2022}\natexlab{}.
\newblock \showarticletitle{One fuzzing strategy to rule them all}. In
  \bibinfo{booktitle}{\emph{Proceedings of the 44th International Conference on
  Software Engineering}}. \bibinfo{pages}{1634--1645}.
\newblock


\bibitem[Xia et~al\mbox{.}(2024)]%
        {xia2024fuzz4all}
\bibfield{author}{\bibinfo{person}{Chunqiu~Steven Xia}, \bibinfo{person}{Matteo
  Paltenghi}, \bibinfo{person}{Jia Le~Tian}, \bibinfo{person}{Michael Pradel},
  {and} \bibinfo{person}{Lingming Zhang}.} \bibinfo{year}{2024}\natexlab{}.
\newblock \showarticletitle{Fuzz4all: Universal fuzzing with large language
  models}. In \bibinfo{booktitle}{\emph{Proceedings of the IEEE/ACM 46th
  International Conference on Software Engineering}}. \bibinfo{pages}{1--13}.
\newblock


\bibitem[Xia et~al\mbox{.}(2025)]%
        {xia2025quantum}
\bibfield{author}{\bibinfo{person}{Shangzhou Xia}, \bibinfo{person}{Jianjun
  Zhao}, \bibinfo{person}{Fuyuan Zhang}, {and} \bibinfo{person}{Xiaoyu Guo}.}
  \bibinfo{year}{2025}\natexlab{}.
\newblock \showarticletitle{Quantum concolic testing}.
\newblock \bibinfo{journal}{\emph{Proceedings of the ACM on Software
  Engineering}} \bibinfo{volume}{2}, \bibinfo{number}{ISSTA}
  (\bibinfo{year}{2025}), \bibinfo{pages}{1146--1166}.
\newblock


\bibitem[Xie et~al\mbox{.}(2025)]%
        {xie2025ase}
\bibfield{author}{\bibinfo{person}{Yuchong Xie}, \bibinfo{person}{Wenhui
  Zhang}, {and} \bibinfo{person}{Dongdong She}.}
  \bibinfo{year}{2025}\natexlab{}.
\newblock \showarticletitle{ZTaint-Havoc: From Havoc mode to zero-execution
  fuzzing-driven taint inference}.
\newblock \bibinfo{journal}{\emph{Proceedings of the ACM on Software
  Engineering}} \bibinfo{volume}{2}, \bibinfo{number}{ISSTA}
  (\bibinfo{year}{2025}), \bibinfo{pages}{917--939}.
\newblock


\bibitem[Yang et~al\mbox{.}(2024)]%
        {yang2024buzzbee}
\bibfield{author}{\bibinfo{person}{Yupeng Yang}, \bibinfo{person}{Yongheng
  Chen}, \bibinfo{person}{Rui Zhong}, \bibinfo{person}{Jizhou Chen}, {and}
  \bibinfo{person}{Wenke Lee}.} \bibinfo{year}{2024}\natexlab{}.
\newblock \showarticletitle{Towards generic database management system
  fuzzing}. In \bibinfo{booktitle}{\emph{33rd USENIX Security Symposium (USENIX
  Security 24)}}. \bibinfo{pages}{901--918}.
\newblock


\bibitem[Ye et~al\mbox{.}(2025)]%
        {ye2025measurement}
\bibfield{author}{\bibinfo{person}{Jiaming Ye}, \bibinfo{person}{Xiongfei Wu},
  \bibinfo{person}{Shangzhou Xia}, \bibinfo{person}{Fuyuan Zhang}, {and}
  \bibinfo{person}{Jianjun Zhao}.} \bibinfo{year}{2025}\natexlab{}.
\newblock \showarticletitle{Is Measurement Enough? Rethinking Output Validation
  in Quantum Program Testing}.
\newblock \bibinfo{journal}{\emph{arXiv preprint arXiv:2509.16595}}
  (\bibinfo{year}{2025}).
\newblock


\bibitem[Ying et~al\mbox{.}(2025)]%
        {ying2025tosem}
\bibfield{author}{\bibinfo{person}{Mingsheng Ying}, \bibinfo{person}{Li Zhou},
  {and} \bibinfo{person}{Gilles Barthe}.} \bibinfo{year}{2025}\natexlab{}.
\newblock \showarticletitle{Laws of Quantum Programming}.
\newblock \bibinfo{journal}{\emph{ACM Transactions on Software Engineering and
  Methodology}} (\bibinfo{year}{2025}).
\newblock


\bibitem[Zhang et~al\mbox{.}(2024b)]%
        {zhang2024effective}
\bibfield{author}{\bibinfo{person}{Cen Zhang}, \bibinfo{person}{Yaowen Zheng},
  \bibinfo{person}{Mingqiang Bai}, \bibinfo{person}{Yeting Li},
  \bibinfo{person}{Wei Ma}, \bibinfo{person}{Xiaofei Xie},
  \bibinfo{person}{Yuekang Li}, \bibinfo{person}{Limin Sun}, {and}
  \bibinfo{person}{Yang Liu}.} \bibinfo{year}{2024}\natexlab{b}.
\newblock \showarticletitle{How effective are they? exploring large language
  model based fuzz driver generation}. In \bibinfo{booktitle}{\emph{Proceedings
  of the 33rd ACM SIGSOFT International Symposium on Software Testing and
  Analysis}}. \bibinfo{pages}{1223--1235}.
\newblock


\bibitem[Zhang et~al\mbox{.}(2025)]%
        {zhang2025waltzz}
\bibfield{author}{\bibinfo{person}{Lingming Zhang}, \bibinfo{person}{Binbin
  Zhao}, \bibinfo{person}{Jiacheng Xu}, \bibinfo{person}{Peiyu Liu},
  \bibinfo{person}{Qinge Xie}, \bibinfo{person}{Yuan Tian},
  \bibinfo{person}{Jianhai Chen}, {and} \bibinfo{person}{Shouling Ji}.}
  \bibinfo{year}{2025}\natexlab{}.
\newblock \showarticletitle{Waltzz:$\{$WebAssembly$\}$ Runtime Fuzzing with
  $\{$Stack-Invariant$\}$ Transformation}. In \bibinfo{booktitle}{\emph{34th
  USENIX Security Symposium (USENIX Security 25)}}.
  \bibinfo{pages}{6159--6178}.
\newblock


\bibitem[Zhang et~al\mbox{.}(2024a)]%
        {zhang2024resolverfuzz}
\bibfield{author}{\bibinfo{person}{Qifan Zhang}, \bibinfo{person}{Xuesong Bai},
  \bibinfo{person}{Xiang Li}, \bibinfo{person}{Haixin Duan},
  \bibinfo{person}{Qi Li}, {and} \bibinfo{person}{Zhou Li}.}
  \bibinfo{year}{2024}\natexlab{a}.
\newblock \showarticletitle{$\{$ResolverFuzz$\}$: Automated Discovery of
  $\{$DNS$\}$ Resolver Vulnerabilities with $\{$Query-Response$\}$ Fuzzing}. In
  \bibinfo{booktitle}{\emph{33rd USENIX Security Symposium (USENIX Security
  24)}}. \bibinfo{pages}{4729--4746}.
\newblock


\bibitem[Zhao(2020)]%
        {zhao2020quantum}
\bibfield{author}{\bibinfo{person}{Jianjun Zhao}.}
  \bibinfo{year}{2020}\natexlab{}.
\newblock \showarticletitle{Quantum software engineering: Landscapes and
  horizons}.
\newblock \bibinfo{journal}{\emph{arXiv preprint arXiv:2007.07047}}
  (\bibinfo{year}{2020}).
\newblock


\bibitem[Zhao et~al\mbox{.}(2023a)]%
        {zhao2023bugs4q}
\bibfield{author}{\bibinfo{person}{Pengzhan Zhao}, \bibinfo{person}{Zhongtao
  Miao}, \bibinfo{person}{Shuhan Lan}, {and} \bibinfo{person}{Jianjun Zhao}.}
  \bibinfo{year}{2023}\natexlab{a}.
\newblock \showarticletitle{Bugs4Q: A benchmark of existing bugs to enable
  controlled testing and debugging studies for quantum programs}.
\newblock \bibinfo{journal}{\emph{Journal of Systems and Software}}
  \bibinfo{volume}{205} (\bibinfo{year}{2023}), \bibinfo{pages}{111805}.
\newblock


\bibitem[Zhao et~al\mbox{.}(2023b)]%
        {zhao2023qchecker}
\bibfield{author}{\bibinfo{person}{Pengzhan Zhao}, \bibinfo{person}{Xiongfei
  Wu}, \bibinfo{person}{Zhuo Li}, {and} \bibinfo{person}{Jianjun Zhao}.}
  \bibinfo{year}{2023}\natexlab{b}.
\newblock \showarticletitle{Qchecker: Detecting bugs in quantum programs via
  static analysis}. In \bibinfo{booktitle}{\emph{2023 IEEE/ACM 4th
  International Workshop on Quantum Software Engineering (Q-SE)}}. IEEE,
  \bibinfo{pages}{50--57}.
\newblock


\bibitem[Zhu et~al\mbox{.}(2022)]%
        {zhu2022csur}
\bibfield{author}{\bibinfo{person}{Xiaogang Zhu}, \bibinfo{person}{Sheng Wen},
  \bibinfo{person}{Seyit Camtepe}, {and} \bibinfo{person}{Yang Xiang}.}
  \bibinfo{year}{2022}\natexlab{}.
\newblock \showarticletitle{Fuzzing: a survey for roadmap}.
\newblock \bibinfo{journal}{\emph{ACM Computing Surveys (CSUR)}}
  \bibinfo{volume}{54}, \bibinfo{number}{11s} (\bibinfo{year}{2022}),
  \bibinfo{pages}{1--36}.
\newblock


\end{thebibliography}
\end{document}